\documentclass[useAMS,usenatbib]{mn2e}

\usepackage{graphicx}

\def\lsim{\mathrel{\rlap{\lower 3pt \hbox{$\sim$}} \raise 2.0pt \hbox{$<$}}}
\def\gsim{\mathrel{\rlap{\lower 3pt \hbox{$\sim$}} \raise 2.0pt \hbox{$>$}}}

\def\kms{{\rm km}\,{\rm s}^{-1}}
\def\ad{\Delta\theta}
\def\pd{{\rm R}_\perp}

\def\pks2155{{\mbox PKS~2155-304}}

\title[The Environment of PKS~2155-304]{The Cluster--Scale Environment of PKS~2155-304\thanks{Based 
on  observations  undertaken  at  the  6.5\,meter Magellan Telescopes.}}
\author[Farina et al.]{
	\parbox{\textwidth}{
	E.~P.~Farina$^{1,}$\thanks{E--mail: {\texttt emanuele.paolo.farina@gmail.com}},
	M.~Fumagalli$^{2}$, 
	R.~Decarli$^{1}$, and
	N.~Fanidakis$^{1}$}\vspace{0.6cm}\\
	\parbox{\textwidth}{
	$^{1}$ Max--Planck--Institut f{\"u}r Astronomie --- K{\"o}nigstuhl 17, D-69117 Heidelberg, Germany\\
	$^{2}$ Institute for Computational Cosmology and Centre for Extragalactic Astronomy, Department of Physics, Durham University --- South Road, Durham, DH1 3LE, UK\\
	}}
\begin{document}

\date{Submitted on \today}

\pagerange{\pageref{firstpage}--\pageref{lastpage}} \pubyear{2015}

\maketitle

\label{firstpage}

\begin{abstract}

\pks2155\ is one of the brightest extragalactic source in the 
X--ray and EUV bands, and is a prototype for the BL~Lac class 
of objects.
In this paper we investigate the large--scale environment 
of this source using new multi--object as well as long--slit 
spectroscopy, together with archival spectra and optical images.
We find clear evidence of a modest overdensity of galaxies at 
$z=0.11610\pm0.00006$, consistent with previous determinations 
of the BL~Lac redshift.
The galaxy group has a radial velocity dispersion of 
250$^{+80}_{-40}$\,km\,s$^{-1}$ and a virial radius
of $0.22$\,Mpc, yielding a role--of--thumb estimate 
of the virial mass of M$_{\rm vir}$$\sim$1.5$\times$10$^{13}$\,M$_\odot$,
i.e., one order of magnitude less than what observed in 
other similar objects. 
This result hints toward a relatively wide diversity in the 
environmental properties of BL~Lac objects.

% Multi--object spectroscopy of the sources in the field confirms early 
% suggestions of the presence of a moderate overdensity of galaxies 
% physically associated with the BL~Lac. 
% We obtain for this galaxy group a redshift z$=$0.11610$\pm$0.00006,
% and a radial velocity dispersion of 250$^{+80}_{-40}$\,km\,s$^{-1}$.
% Under the hypothesis that the distribution of the galaxies is a 
% good proxy for the underlying mass distribution we estimate a 
% virial mass of . 

\end{abstract}

\begin{keywords}
\mbox{BL Lacertae objects:\,individual:\,\pks2155}; \mbox{galaxies:\,groups:\,general}
\end{keywords}

\section{Introduction}

BL Lac objects are a subclass of active galactic nuclei (AGN) showing 
a strong, non--thermal variable emission from radio to TeV energies. 
These properties are usually ascribed to the relativistic jet emission 
that is closely aligned with the line--of--sight \citep[][]{Blandford1978}. 
Four decades of studies of the host galaxies and of the BL~Lacs close 
environment have lead to a general consensus that they are mainly 
hosted by luminous elliptical galaxies embedded in small clusters or 
group of galaxies \citep[e.g.][see also \citealt{Falomo2014} for a 
recent review]{Weistrop1979, Wurtz1993, Wurtz1997, Falomo1990, 
Smith1995, Falomo1996, Falomo1999, Scarpa2000, Urry1993, Urry2000, 
Nilsson2008, Kotilainen2011}.
Most of these studies are, however, based on photometric data only, 
and evidence for a rich environment is inferred from the increase 
number counts of sources in the proximity of the BL~Lacs.
Only for a handful of objects the physical association of nearby 
sources has been further confirmed via dedicated spectroscopy 
\citep[e.g.][]{Pesce1994, Pesce1995, Falomo1995, Lietzen2008, Muriel2015}. 

\begin{figure*}
\centering
\includegraphics[width=1.99\columnwidth]{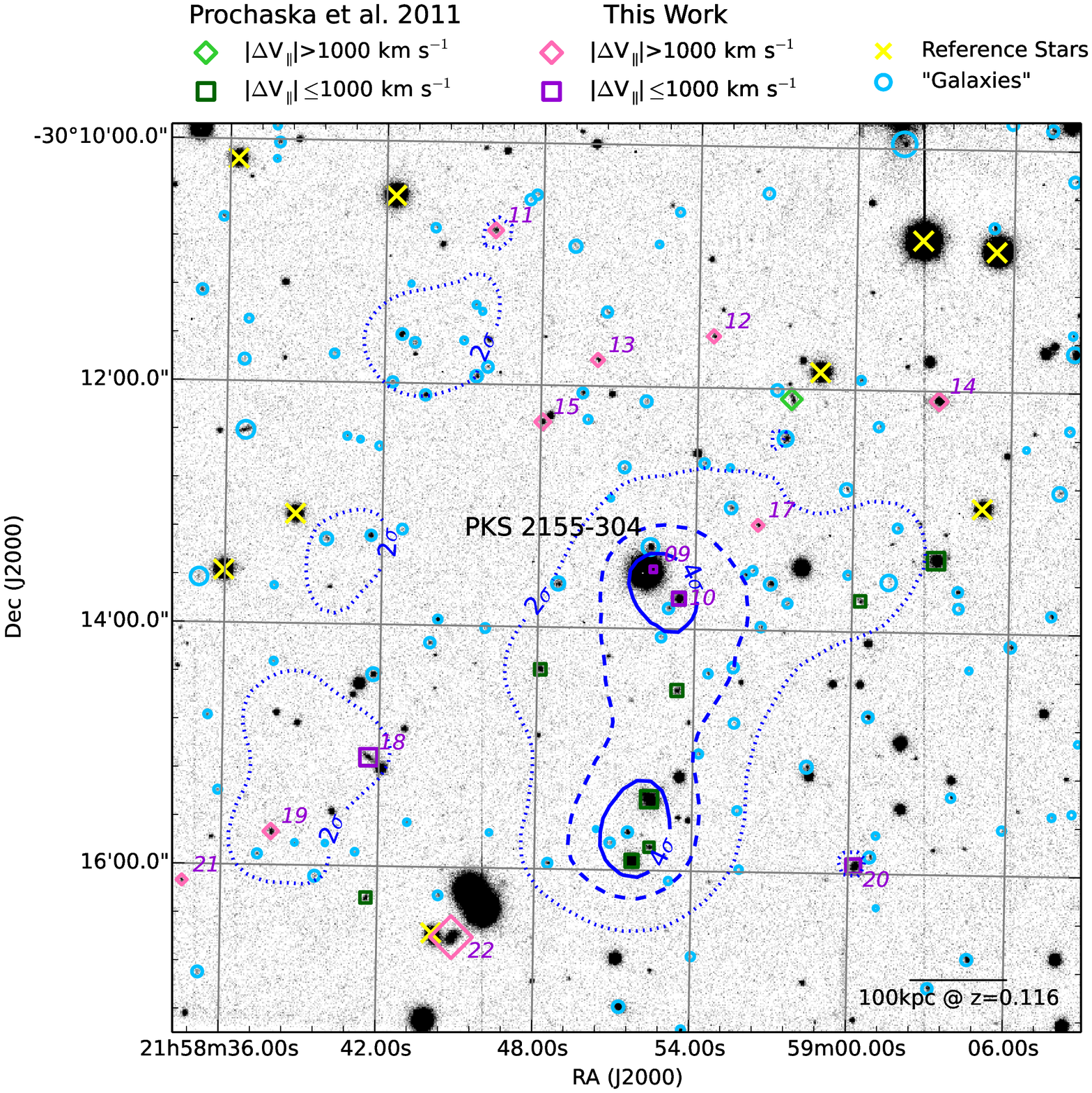}
\caption{
Field of \pks2155\ as imaged in V--band with EMMI \citep[][North is 
up and East is right]{Falomo1993}.
Light blue circles mark objects selected as galaxies on the basis of 
their \texttt{CLASS\_STAR} parameter and not confirmed via dedicated
spectroscopy. 
Purple squares and pink diamonds are the targets of multi--object 
spectroscopy with IMACS.
Dark green squares and light green diamonds are archival data 
from \citet{Prochaska2011}.
The different shape of the points indicates whether a source is 
located within $\pm$1000\,$\kms$ from the average redshift of 
the overdensity (squares) or not (diamonds).
The size of the points scales with the effective radius of the 
objects.
Blue dotted, dashed, and solid lines are the 2, 3, and 4$\sigma$ 
probability level to have an overdensity of galaxy at z$=$0.116
estimated using the CRS method (see Sec.\,\ref{sec:over}).
The nine yellow crosses show the position of the reference stars 
used for mask alignment.
}\label{fig:ntt}
\end{figure*}

In this paper we investigate the close galactic environment of 
\pks2155, often considered an archetypal of X--ray selected 
BL~Lac objects \citep{Schwartz1979, Griffiths1979}.
It is one of the most luminous, non transient, extragalactic source 
known and it shows a rapid and strong variability in the whole 
electromagnetic spectrum \citep[e.g.,][]{Smith1992, Sembay1993, 
Fan2000, Aharonian2007, Rieger2010, Zhang2014, Sandrinelli2014a, 
Sandrinelli2014b}. 
The host of \pks2155\ is a luminous elliptical galaxy with 
M$_{\rm R}$=-24.4 \citep{Falomo1991, Kotilainen1998} at redshift 
z=0.116 (\citealt{Falomo1993}; \citealt{Sbarufatti2006}\footnote{
Spectra available in the ZBLLac archive:\newline 
\texttt{http://archive.oapd.inaf.it/zbllac/}}). 
Optical imaging of the field of \pks2155\ revealed the presence 
of a moderated overdensity of galaxies \citep{Falomo1991, Falomo1993,
Wurtz1997}. 
This finding is corroborated by the detection in high--resolution NIR 
images of 5~galaxies located within $\sim$20\arcsec\ ($\sim$40\,kpc at 
z=0.116) from \pks2155 \citep{Liuzzo2013} and by spectroscopic observations 
of some of the sources \citep{Falomo1991, Falomo1993}.
We here present a comprehensive study of the field of \pks2155\ 
using new spectra gathered with the Magellan Telescope and archival 
data from \citet{Prochaska2011}. 
This unique dataset, including $\gsim$34\% of all the galaxies brighter
than R=20\,mag located within 4\arcmin\ from the BL~Lac (see Sec.\,\ref{sec:res}), 
will allow us to probe the galactic environment with unprecedent details.

Throughout this paper we assume a concordance cosmology with 
H$_0$=70\,km\,s$^{-1}$\,Mpc$^{-1}$, $\Omega_{\rm m}$=0.3, and 
\mbox{$\Omega_\Lambda$=0.7}. In this cosmology, at z=0.116 
an angular scale of $\ad$$=$1\arcsec\ corresponds to a 
proper transverse separation of $\pd$$=$2.1\,kpc.
All the quoted magnitudes are expressed in the AB standard 
photometric system \citep{Oke1974,Oke1983}.

\section{Target Selection, Observations and Data Analysis}

In this Section we detail the selection of the targets and 
we present the multi--object and long--slit spectroscopic 
observations performed with the Inamori--Magellan Areal 
Camera \& Spectrograph \citep[IMACS,][]{Dressler2011} 
mounted on the 6.5m Magellan Telescope Baade (Las Campanas 
Observatory, Chile).
The data reduction process and the analysis of the spectra
is also described.
% We also summarise the data reduction process and the 
% analysis of the spectra.
Eventually, we present a summary of the archival photometric 
and spectroscopic data from \citet{Prochaska2011} used in 
this work.

\subsection{Multi--Object Spectroscopy}\label{sec:mos}

Targets for the multi--object spectroscopy were designated from the 
analysis of the broad V--band image collected by \citet{Falomo1993} 
with the ESO Multi--Mode Instrument \citep[EMMI,][]{Dekker1986} 
on the ESO New Technology Telescope (NTT, Fig.\,\ref{fig:ntt}).
The size of this image is 7\farcm5$\times$7\farcm5. This roughly 
corresponds to 950$\times$950\,kpc$^2$ at z=0.116, and well fits
inside the f/4 camera field--of--view of IMACS (15\farcm4$\times$15\farcm4).
Data were recalibrated by cross--matching the sources present in the
field with the NOMAD catalogue \citep{Zacharias2004, Zacharias2005},
reaching an uncertainty on the photometric zero point of $\sim$0.1\,mag.
In order to achieve astrometric accuracy better than 0\farcs5 on the 
whole frame (required for mask alignment), the astrometric solution 
was refined with the \texttt{ASTROMETRY.NET} software \citep{Lang2010}.
During the imaging observations the seeing was 1\farcs6 and the reached 
5$\sigma$ detection limit (estimated from the rms of the sky counts 
integrated within a seeing radius) was V$_{\rm lim}$$\approx$22.8\,mag.

\begin{figure}
\centering
\includegraphics[width=0.99\columnwidth]{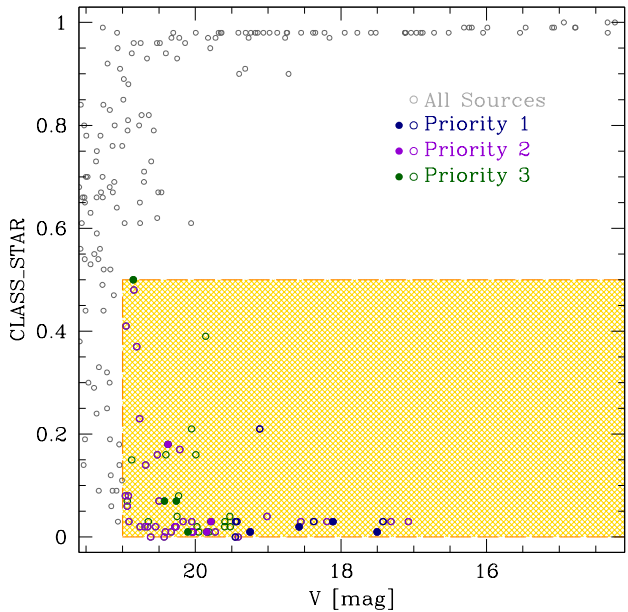}
\caption{\texttt{SEXTRACTOR} star/galaxy classifier (\texttt{CLASS\_STAR})
as a function of the apparent V--band magnitude of the sources detected in
the NTT image. Objects selected for multi--object spectroscopy lie within 
the yellow hashed region. Different colours of the circles correspond to the 
priority flag assigned to each target: blue, purple, and green for
priority 1, 2, and 3, respectively (see Sec.\,\ref{sec:mos} for details).
Filled circles are objects spectroscopically observed with IMACS.
}\label{fig:cs}
\end{figure}

We used \texttt{SEXTRACTOR} \citep{Bertin1996} to identify and classify
sources. 
An object was considered as possible target for spectroscopy if its 
\texttt{SEXTRACTOR} star/galaxy classifier (the \texttt{CLASS\_STAR} 
parameter) was smaller than 0.5 (i.e., more likely to be an extended 
source) and it was brighter than V$=$21\,mag.
In Figure~\ref{fig:cs}, we plot the \texttt{CLASS\_STAR} classifier
as a function of the V--band magnitude. 
The adopted cut in magnitude allows us to avoid faint sources that 
\texttt{SEXTRACTOR} may fail to classify due their low signal--to--noise 
in the image. 
The conservative limit for the \texttt{CLASS\_STAR} parameter was chosen 
to remove the most obvious stars, yet to not miss possible compact galaxies. 
However, we observe that the majority of the selected sources are in the 
\texttt{CLASS\_STAR}$\sim$0 regime.
The IMACS Mask Generation Software (\texttt{maskgen}\footnote{\texttt{http://code.obs.carnegiescience.edu/maskgen}.})
requires a priority flag to solve conflict between possible overlapping 
spectra and to maximise the number of sources in the mask. 
We thus arbitrarily assigned priority~1 (high probability to be included 
in the mask) to objects with V$<$19.5\,mag and ellipticity $e$$>$0.2, 
priority~2 to those sources with either V$<$19.5\,mag or $e$$>$0.2 but 
not both, and priority~3 to the remaining sources (see Fig.\,\ref{fig:cs}). 
This corresponds to prioritise for bright elliptical galaxies, that are
expected to be the dominant population of a putative low redshift cluster.
In addition to this list of targets we considered also the four galaxies 
G1, G2, G5, and G6 (following the labels assigned by \citealt{Falomo1993} 
and \citealt{Liuzzo2013}) observed in high resolution NIR images by 
\citet{Liuzzo2013}. 
These sources are positioned at less than $\sim$20\arcsec\ from \pks2155\ 
and we assigned them a priority of~1. 
Eventually, sources from the sample of \citet{Prochaska2011} located outside
the NTT image field--of--view were included with priority of~3.
To avoid the gaps between the 8~chips of the CCD mosaic (7\farcm7$\times$3\farcm8 
each), we centred our frame 1\farcm5~North and 1\farcm5~West from the
position of \pks2155\ and the mask position angle was set to 270\degr.
Sixteen apertures of 1\arcsec$\times$12\arcsec were cut into the mask,
corresponding to the targets listed in Table~\ref{tab:sample} and 
showed in Figure\,\ref{fig:ntt}.
In addition, nine reference stars were used for the mask alignment
(see Fig.\,\ref{fig:ntt}).

Multi--object spectroscopy data were gathered on October 5th, 2013 using 
the IMACS grating 300-l. 
By selecting a central wavelength of 6000\,\AA\ we were able to almost 
continuously cover the spectral region from $\sim$3500\,\AA\ to 
$\sim$9000\,\AA\ (with small variations due to the different position
of the sources). 
Three contiguous exposures of 1200\,s each were collected in clear sky 
conditions and with an average seeing of $\sim$0\farcs7.
The \texttt{COSMOS} package\footnote{\texttt{http://code.obs.carnegiescience.edu/cosmos},
see also \citet{Kelson2003} for details on the sky subtraction.} was 
employed for the data reduction. 
Flux calibration was performed using a spectrum of the standard star 
Feige~110 acquired during the same night and rescaling the spectra to 
the V--band magnitude of the targets \citep[see][for further details]{Decarli2008}. 
Galactic extinction was accounted for according to the map of dust 
reddening from \citet[][i.e., E(B-V)$=$0.048]{Schlafly2014}.
The reduced spectra typically have a signal--to--noise ratio 
S/N$>$3~per pixel at $\lambda$=6000\AA\ and are showed in 
Figure~\ref{fig:mos} in the online edition of the Journal.

To determine the source redshifts we followed a three~step procedure.
First we find the redshift (in step of $\Delta$z=0.1) that best 
match each spectrum with the \citet{Kinney1996} galaxy templates 
rescaled to the magnitude of the target.
Then the redshift was refined measuring the position of prominent 
absorption and emission features and matching them with the line 
list from the SDSS data set \citep[e.g.][]{Oh2011}.
Finally, spectra were independently visually inspected by the 
authors to verify their redshifts and a quality flag was assigned:
$Q_{\rm z}$=1 means reliable redshift from apparent emission and/or 
absorption features; and $Q_{\rm z}$=2 means probable redshift.
With the exception of PKS~2155-304\_22, that seems to show the 
presence of rest--frame Balmer lines\footnote{We notice that 
this contaminant lies just at the edge of our selection criteria (see 
Sec.\,\ref{sec:mos}), with V$=$20.85 and \texttt{CLASS\_STAR}$=$0.5.}, 
a redshift was established for all the sources from the measure of 
two or more features (see Tab.\,\ref{tab:sample}).
The uncertainties on the redshifts are typically $\sigma_{\rm z}$$\sim$0.0001.

\begin{table*}
\centering
\caption{
List of the targets observed with multi--object spectroscopy: 
our identification label of the object (ID); position (RA, Dec);
B, V, and R--band magnitude derived from the NTT image and from 
\citet[][B, V, R]{Prochaska2011}; angular separation from \pks2155\ 
($\Delta\theta$); redshift (z); visual inspection quality flag of
the determined redshift ($Q_{\rm z}$, see text for details); and 
alternative label from \citet{Falomo1993} and \citet[][Alt. ID]{Prochaska2011}.  
}\label{tab:sample}
\begin{tabular}{lccccccccl}
\hline
ID               & RA          & Dec         &  B             & V              & R	       & $\Delta\theta$   & z	              & $Q_{\rm z}$ & Alt. ID  		           \\
                 & (J2000)     & (J2000)     &  (mag)         & (mag)          & (mag)	       & (arcsec)	  &                   &	            &  			           \\
\hline
PKS~2155-304\_09 & 21:58:52.38 & -30:13:30.5 & \dots	      & \dots	       & \dots  	& \phantom{10}4.3 & 0.1167$\pm$0.0001 & 1	    & G1$^{\rm a}$		   \\
PKS~2155-304\_10 & 21:58:53.33 & -30:13:44.6 & 19.93$\pm$0.10 & 18.55$\pm$0.07 & 18.29$\pm$0.06 & \phantom{1}21.1 & 0.1166$\pm$0.0001 & 1	    & G2$^{\rm a}$, 1790$^{\rm b}$ \\ 
PKS~2155-304\_11 & 21:58:46.16 & -30:10:43.6 & 21.32$\pm$0.15 & 20.10$\pm$0.13 & 19.81$\pm$0.07 &	    185.1 & 0.3233$\pm$0.0002 & 2	    & 2085$^{\rm b}$		   \\
PKS~2155-304\_12 & 21:58:54.56 & -30:11:34.6 & 21.59$\pm$0.15 & 20.38$\pm$0.31 & 20.21$\pm$0.07 &	    122.0 & 0.2482$\pm$0.0003 & 2	    & 1760$^{\rm b}$		   \\
PKS~2155-304\_13 & 21:58:50.13 & -30:11:47.1 & 22.21$\pm$0.27 & 20.26$\pm$0.16 & 19.76$\pm$0.07 &	    108.0 & 0.3119$\pm$0.0001 & 1	    & 1928$^{\rm b}$		   \\
PKS~2155-304\_14 & 21:59:03.20 & -30:12:05.0 & 19.29$\pm$0.10 & 18.11$\pm$0.33 & 17.74$\pm$0.06 &	    168.5 & 0.1202$\pm$0.0001 & 1	    & 1389$^{\rm b}$		   \\
PKS~2155-304\_15 & 21:58:48.05 & -30:12:18.3 & 21.65$\pm$0.21 & 19.78$\pm$0.22 & 19.14$\pm$0.07 & \phantom{1}90.3 & 0.3154$\pm$0.0001 & 1	    & 2158$^{\rm b}$		   \\
PKS~2155-304\_17 & 21:58:56.36 & -30:13:07.8 & 22.00$\pm$0.23 & 20.42$\pm$0.05 & 19.59$\pm$0.07 & \phantom{1}60.5 & 0.3127$\pm$0.0001 & 1	    & 1689$^{\rm b}$		   \\
PKS~2155-304\_18 & 21:58:41.59 & -30:15:05.8 & 20.75$\pm$0.14 & 19.25$\pm$0.63 & 18.92$\pm$0.07 &	    165.1 & 0.1151$\pm$0.0003 & 2	    & 2210$^{\rm b}$		   \\
PKS~2155-304\_19 & 21:58:37.91 & -30:15:43.5 & 21.04$\pm$0.14 & 19.84$\pm$0.30 & 19.36$\pm$0.07 &	    225.9 & 0.2568$\pm$0.0001 & 1	    & 2364$^{\rm b}$		   \\
PKS~2155-304\_20 & 21:59:00.21 & -30:15:56.2 & 19.87$\pm$0.11 & 18.57$\pm$0.21 & 18.23$\pm$0.06 &	    178.5 & 0.1157$\pm$0.0001 & 1	    & 1496$^{\rm b}$		   \\
PKS~2155-304\_21 & 21:58:34.53 & -30:16:08.1 & 21.80$\pm$0.15 & 20.85$\pm$0.06 & 20.98$\pm$0.09 &	    275.8 & \dots             & 2	    & 2499$^{\rm b}$		   \\
PKS~2155-304\_22 & 21:58:44.87 & -30:16:34.5 & \dots	      & 17.50$\pm$0.52 & \dots  	&	    205.1 & 0.1058$\pm$0.0003 & 2	    & \dots			   \\
PKS~2155-304\_26 & 21:59:31.12 & -30:16:15.0 & 18.28$\pm$0.09 & \dots	       & 16.54$\pm$0.06 &	    531.3 & 0.1154$\pm$0.0001 & 1	    &  351$^{\rm b}$		   \\
PKS~2155-304\_28 & 21:59:12.70 & -30:11:32.0 & 19.26$\pm$0.09 & \dots	       & 18.20$\pm$0.06 &	    292.9 & 0.1485$\pm$0.0001 & 1	    & 1027$^{\rm b}$		   \\
PKS~2155-304\_29 & 21:58:27.60 & -30:14:16.0 & 21.53$\pm$0.20 & \dots	       & 19.15$\pm$0.07 &	    320.3 & 0.3137$\pm$0.0005 & 2	    & 2762$^{\rm b}$		   \\
\hline																											  
\multicolumn{9}{l}{$^{\rm a}$ \citet{Falomo1993}.   } \\
\multicolumn{9}{l}{$^{\rm b}$ \citet{Prochaska2011}.} \\
\end{tabular}
\end{table*}

\subsection{Long--Slit Spectroscopy}

\begin{figure}
\centering
\includegraphics[width=0.99\columnwidth]{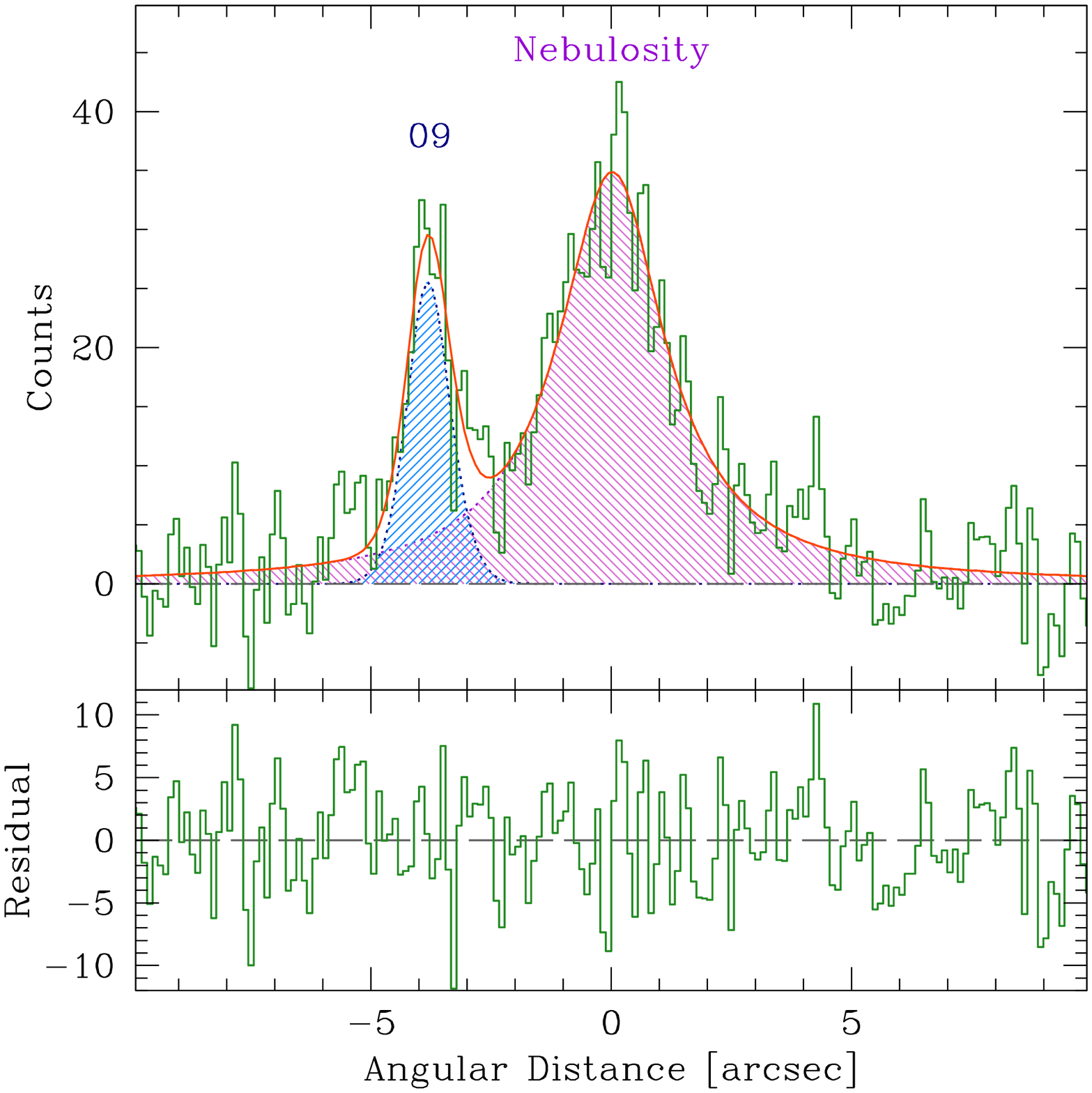}
\caption{
Example of our deblending procedure on a 3~pixel slice 
of the 2D long--slit spectrum centred at $\lambda$=5500\AA.\newline
{\it Top Panel} ---  
Spatial profile of the two spectra (green line). 
The contributions of PKS~2155-304\_09 and of the \pks2155's 
nebulosity were deblended performing a simultaneous 
fit with a Gaussian and a Lorentzian function, 
respectively (blue and purple shaded areas). 
The result of the fit is also shown (orange line).\newline
{\it Bottom Panel} --- Residual spectrum.
}\label{fig:deb}
\end{figure}

In addition to multi--object spectroscopy, we acquired a long--slit 
spectrum of the two closest galaxies to \pks2155\ \citep[i.e., G1 
and G5;][]{Falomo1993, Liuzzo2013}.
Data were gathered during the same night of the multi--object spectroscopy
with a seeing of $\sim$0\farcs5.
Two subsequent exposures of 1500\,s each were acquired with the 0\farcs7 
long--slit and the 8\fdg6 blaze angle of the IMACS grating \mbox{600-l}, 
allowing
us to nominally cover the wavelength range between 3650\AA\ and 6750\AA.
The position angle was set to 104\degr\ in order to simultaneously collect 
spectra of both the galaxies and of the BL~Lac nebulosity.

Standard \texttt{IRAF}\footnote{\texttt{IRAF} \citep{Tody1986, Tody1993}, 
is distributed by the National Optical Astronomy Observatories, which are 
operated by the Association of Universities for Research in Astronomy, Inc., 
under cooperative agreement with the National Science Foundation.} 
procedures for optical spectroscopy were adopted in the data reduction.
First, each of the 8 detectors of mosaic CCD camera was bias--subtracted 
and flat--fielded independently.
Single exposures were then aligned, combined, and cleaned for cosmic--rays 
using the Laplacian edge detection algorithm presented by \citet{vanDokkum2001}.
The wavelength calibration was applied taking as reference the spectrum of 
He+Ne+Ar arc lamps.

The small angular separation of PKS~2155-304\_09 from \pks2155
causes its spectrum to be embedded within the \pks2155\ nebulosity 
(see Fig.\,\ref{fig:deb}). 
In order to disentangle the two contributions we make use of our own 
\texttt{python} routine. Namely, first the sky emission was removed 
considering regions free from emission by astrophysical sources. 
Then, at each pixel of the dispersion axis we sliced the spectra along 
the spatial direction and we fitted the profile with a combination of 
a Gaussian and a Lorentzian functions to model the contributions of 
PKS~2155-304\_09 and of the nebulosity, respectively  (see Fig.\,\ref{fig:deb}).
The area subtended by each of the curves was considered as the flux 
of the sources at that wavelength. The resulting spectra were corrected 
for Galactic extinction and the achieved signal--to--noise ratios per pixel 
are typically S/N$>$4 at $\lambda$=5500\AA\ (see Figures~\ref{fig:lon}
and~\ref{fig:neb} in the online version of MNRAS).

The redshifts of the two galaxies were inferred using the method
described above (see Tab.\,\ref{tab:sampleLS}), while for the 
BL~Lac object we have first normalised to spectrum with a power--law, 
and than searched for the presence of significant lines \citep[i.e., 
with an equivalent width greater than EW$_{\rm min}$=1.4\AA, as 
estimated using the method of][]{Sbarufatti2005a}.
Given this limit, the spectrum does not show the presence of any 
spectral features.
We observe, however, that the redshifted Ca\,{\sc II}~$\lambda\lambda$3934,3968 
and Na\,{\sc D}~$\lambda$5892 lines may be tentatively 
($\sim$2\,$\sigma$) detected if the redshift z=0.116 \citep[][]{Falomo1993, 
Sbarufatti2006} is assumed (see Fig.\,\ref{fig:neb} in the electronic
version of the Journal).
Higher signal--to--noise ratio spectra are mandatory to confirm 
the presence of these features.
We point out that these absorptions could also arise from the cool gas in 
the circum--galactic medium of the galaxy group \citep[e.g.][]{Boksenberg1978, 
Blades1981, Richter2011}. 
In this case, the estimates of \citet{Falomo1993} and \citet{Sbarufatti2006}
have to be considered as a lower limit for the redshift of \pks2155.
This latter scenario is discouraged by the identification of the bright 
elliptical host galaxy \citep{Falomo1996, Kotilainen1998} that could be 
used as a {\it standard candle} to determine the redshift \citep[e.g.][]{
Sbarufatti2005b}, and by the observation a strong emission of \pks2155\ 
in the TeV regime \citep{Chadwick1999a, Chadwick1999b, Aharonian2005} that
push the redshift at z$\ll$1 due to the interaction of $\gamma$--ray photons
with the lower frequency photons of the extragalactic background light
\citep[see e.g.][]{Aharonian2013}.

% We determined the redshift of all the three sources, and the
% derived values for PKS~2155-304\_09 and for \pks2155\ are 
% consistent with early estimate by \citet{Falomo1993} and 
% \citet[][see Table~\ref{tab:sampleLS}]{Sbarufatti2006}.

\begin{table*}
\centering
\caption{
List of targets observed with long--slit spectroscopy: our 
identification label of the object (ID); position (RA, Dec); 
angular separation from \pks2155\ ($\Delta\theta$); redshift 
(z); visual inspection quality flag of the determined redshift 
($Q_{\rm z}$); and alternative label from \citet{Falomo1993} 
and \citet[][Alt. ID]{Liuzzo2013}.
}\label{tab:sampleLS}
\begin{tabular}{lcccccc}
\hline
ID                & RA          & Dec         & $\Delta\theta$ & z      & $Q_{\rm z}$ & Alt. ID      \\
                  & (J2000)     & (J2000)     & (arcsec)       &        & 	      & 	     \\
\hline
PKS~2155-304\_08  & 21:58:52.63 & -30:13:31.4 &  7.3           & 0.2008$\pm$0.0005 & 2	      & G5$^{\rm a}$ \\
PKS~2155-304\_09  & 21:58:52.38 & -30:13:30.5 &  4.3           & 0.1168$\pm$0.0003 & 1	      & G1$^{\rm b}$ \\
PKS~2155-304\_Neb & 21:58:52.13 & -30:13:29.6 &  2.6           & \dots             & \dots    & \dots	     \\ 
\hline																											  
\multicolumn{6}{l}{$^{\rm a}$ \citet{Liuzzo2013}.} \\
\multicolumn{6}{l}{$^{\rm b}$ \citet{Falomo1993}.} \\
\end{tabular}
\end{table*}

\subsection{Archival Data}

To investigate the environment of \pks2155 we also take advantage 
of the study of \citet{Prochaska2011} aimed to identify galaxies 
within $\sim$1\,Mpc from ultraviolet bright quasars.
In summary: 
a field of 22\arcmin$\times$22\arcmin\ (i.e., 2.7$\times$2.7\,Mpc$^2$ 
at z=0.116) centred on \pks2155\ was imaged in B-- and R--band with 
the Swope 40\arcsec\ telescope in photometric conditions reaching 
magnitude limits of B$_{\rm lim}$$\approx$24.0\,mag and 
R$_{\rm lim}$$\approx$22.6\,mag.
In addition, multi--object spectroscopy of 160 spatially extended 
sources brighter than R=20\,mag was collected using the WFCCD 
instrument on the Dupont 100\arcsec\ telescope  \citep[see][for 
further details]{Prochaska2011}.
We observe that, in the overlapping region between the NTT V--band 
image and the B-- and R--band images from \citet{Prochaska2011},
there is a discrepancy on the photometric classification of the 
sources:
employing their cut on the \texttt{SEXTRACTOR} star/galaxy classifier 
(i.e., \texttt{CLASS\_STAR}$<$0.98), $\sim$30\% of the sources we 
photometrically selected for our multi--object spectroscopic observations 
(see Sec.\,\ref{sec:mos}) are not recognised as ``galaxy'' by 
\citet{Prochaska2011}.
This is most probably due to the different atmospheric conditions and 
to subtle difference in the parameters ingested to \texttt{SEXTRACTOR} 
to detect sources in the images.
We obtained spectra of four sources that in \citet{Prochaska2011} have 
\texttt{CLASS\_STAR}$\geq$0.98 (i.e., \pks2155\_12, \pks2155\_13, 
\pks2155\_17, and \pks2155\_21), and three of those turn out to be 
spectroscopically confirmed galaxies.
In order to estimate the overdensity of sources around \pks2155
(see Sec.\,\ref{sec:over}) we thus decided to slightly relax 
the constrain on \texttt{CLASS\_STAR} in \citet{Prochaska2011}.
In the following we will consider as {\it galaxies} (based only on the 
photometric information) objects present in the \citet{Prochaska2011} 
sample that have \texttt{CLASS\_STAR}$<$1.
This will allow us to recover the missing fraction of galaxies,
and, at the same time, to take advantage of the wider sky region
covered by the B-- and R--band images.

% 
% We observe that \citet{Prochaska2011} do not classify as galaxy 
% (on the basis of the \texttt{SEXTRACTOR} parameter \texttt{CLASS\_STAR}) 
% $\sim$30\% of the sources we selected for our multi--object 
% spectroscopic observations.
% This is most probably due to the different atmospheric conditions and 
% to subtle difference in the parameters ingested to \texttt{SEXTRACTOR} 
% to detect sources in the images.
% To recover this missing fraction we slightly relaxed the constrain on 
% \texttt{CLASS\_STAR}, and in the following we will consider as 
% {\it galaxies} sources present in the \citet{Prochaska2011} sample 
% that have \texttt{CLASS\_STAR}$<$1.
% 

\section{Results}\label{sec:res}

\begin{figure}
\centering
\includegraphics[width=1.00\columnwidth]{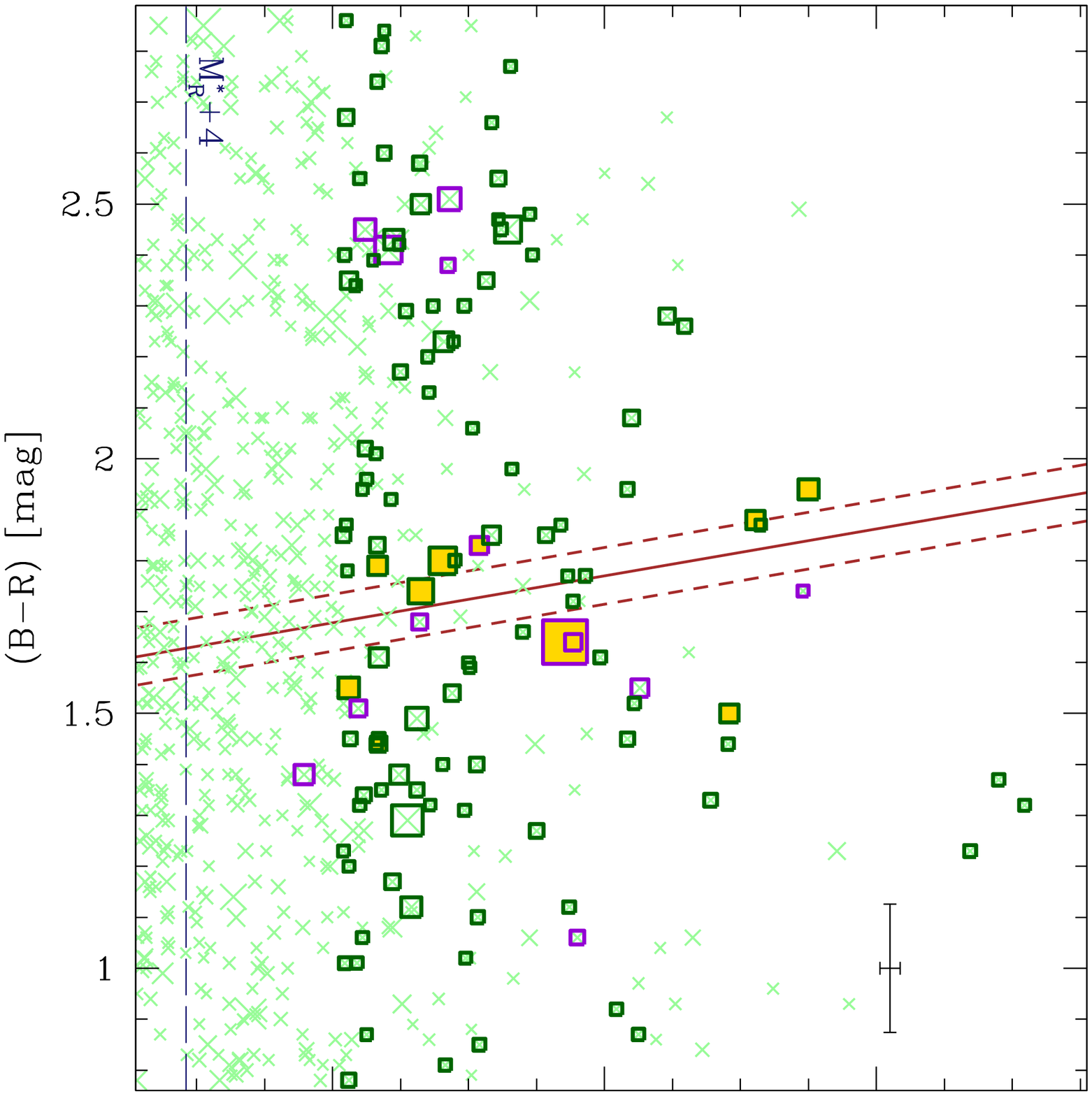}
\includegraphics[width=1.00\columnwidth]{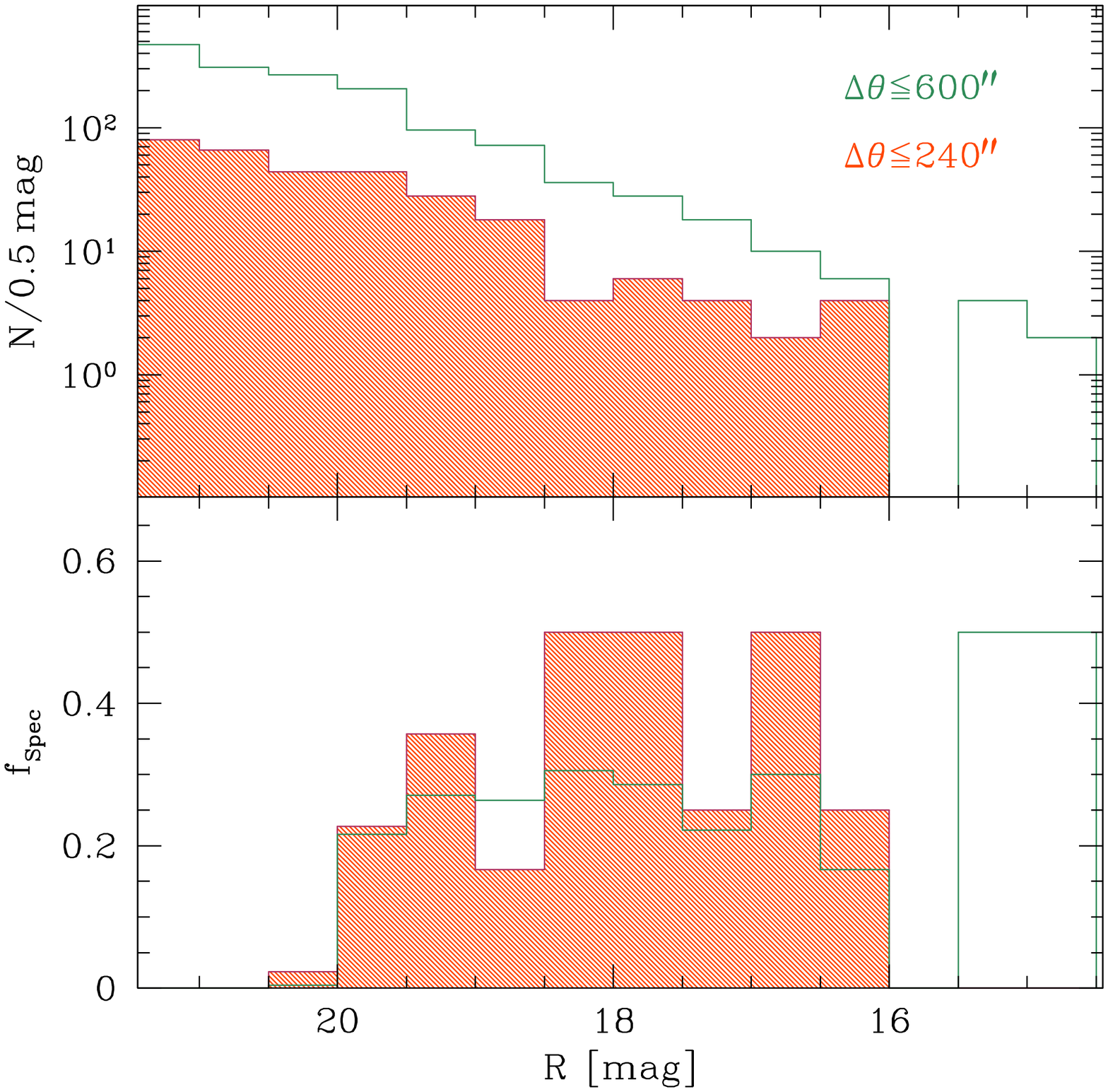}
\caption{
Colour--magnitude diagram of the sources in the 
field of \pks2155. Galaxies \pks2155\_08, \pks2155\_09, 
and the BL~Lac host are not detected in the BVR images,
and are not showed.
\newline
{\it Top Panel} ---  
Pale green crosses indicate the position of galaxies present in 
the \citet{Prochaska2011} sample with
a size that is proportional to the distance from \pks2155.
Framed crosses are targets for which a redshift was determined
from our (purple) and \citet[][dark green]{Prochaska2011} multi--object
spectroscopy. Points filled with yellow
are targets within 240\arcsec\ and 1000\,km\,s$^{-1}$ from 
the BL~Lac object.
% The locus of early (pink circles) and late (light blue circles) 
% type galaxies member of the Coma cluster shifted at the redshift 
% of \pks2155\ is also showed \citep[data from \texttt{GOLDMine},][]{Gavazzi2003}
The best fit of Coma cluster red--sequence from \citet{Lopez2004}
shifted at the redshift of \pks2155\ is also showed (see text for
details).
The cross in the bottom right represents typical error bars.\newline
{\it Middle Panel} ---  
Number of galaxies per magnitude bin located within
600\arcsec\ (i.e., $\sim$1.3\,Mpc at z=0.116, green histogram) 
and 240\arcsec\ (500\,kpc, orange shaded histogram) from \pks2155.\newline
{\it Bottom Panel} ---  
Fraction of the sources per magnitude bin for which a redshift
is determined spectroscopically. The colour code is the same of 
the {\it Middle Panel}.
}\label{fig:cm}
\end{figure}

Using IMACS multi--object and long--slit spectroscopy we obtained the 
redshifts of 17 targets in the field of \pks2155. 
With the exception of \pks2155\_22, that possibly presents rest--frame 
Balmer lines, all the sources are located in the 0.106$\lsim$z$\lsim$0.323 
redshift range (see Tables~\ref{tab:sample} and \ref{tab:sampleLS}).
Combining these new data with the sample of \citet{Prochaska2011}
we reach a completeness in spectroscopy of $\sim$34\% ($\sim$25\%) 
for galaxies brighter than R=20\,mag lying within 240\arcsec\ 
(600\arcsec) from the BL~Lac (see Fig.\,\ref{fig:cm}).
We stress that, among the sources observed with multi--object 
spectroscopy, only \pks2155\_09 was chosen with {\it a priori} knowledge 
of the redshift. These two samples seem thus well suited for an unbiased 
study of the galactic environment of \pks2155. 

In the following, we will first investigate the environment of \pks2155\ 
using the photometric information only (Sec.\,\ref{sec:over}). Afterward, 
we will include the spectroscopic information in our analysis (Sec.\,\ref{sec:overspec}).

\subsection{The overdensity around \pks2155}\label{sec:over}

To verify early suggestions for an overdensity of sources around 
\pks2155 \citep{Falomo1991,Falomo1993}, we first measured the 
density of galaxies brighter than R=21.08\,mag in different boxes 
of 5\arcmin$\times$5\arcmin\ spread over the field.
At the redshift of the BL~Lac, this correspond to cut the
luminosity function at M$^\star_{\rm R}$+4 \citep[where 
M$^\star_{\rm R}$ is the characteristic luminosity of galaxies 
derived from][]{Blanton2001}.  
We obtain an average number density of sources of: 
n$_{\rm bkg}$=(1.51$\pm$0.07)\,arcmin$^{-2}$ that is lower than 
what observed in the box centred on the BL~Lac object: 
n$_{\rm BL\,Lac}$=(2.4$\pm$0.3)\,arcmin$^{-2}$, corresponding to 
a 3.3$\sigma$ overdensity. 
In this estimate, galaxies not present in the BVR images \citep[i.e., 
objects detected only in high--resolution NIR images by][]{Liuzzo2013}
were excluded barring the host galaxy of \pks2155.

Since the BVR images considered in this work cover the $4000$\,\AA\ 
break at z$\sim$0.116, we can refine the position and the extent of the 
overdensity using the Cluster--Red--Sequence method \citep[CRS,][]{Gladders2000}.
In summary, we took as reference the red--sequence of the Coma cluster 
\citep[data from the \texttt{GOLDMine} 
archive\footnote{\texttt{http://goldmine.mib.infn.it/}},][]{Gavazzi2003}
shifted at z=0.116 using the \citet{Kinney1996} galaxy templates 
for K-- and filter--corrections (see Fig.\,\ref{fig:cm}).
To each galaxy brighter than M$^\star_{\rm R}$+4 we then assigned 
a probability to be consistent with the Coma red--sequence on the 
basis of their B-R colours and R--band magnitudes. 
Eventually, a luminosity weighted probability density was computed 
using a fixed--kernel smoothing with radius $r_{\rm K}$=60\arcsec. 
In this map \pks2155\ is embedded in a $\gsim$4$\sigma$ 
overdensity (see Fig.\,\ref{fig:ntt}), confirming the 
evidence that a group of galaxies is associated with the 
BL~Lac.

In order to compare these results with previous works, we measured 
the galaxy--BL~Lac angular cross correlation function ($A_{gb}$) 
and spatial covariance function \citep[$B_{gb}$,][]{Longair1979}
of galaxies brighter that M$^\star_{\rm R}$+2.
This cut in the luminosity function was performed to minimise the
uncertainties in $B_{gb}$ \citep[e.g.][]{Yee1999} and for consistency 
with \citet{Wurtz1997}.
We obtain $A_{gb}$=(0.0013$\pm$0.0014)\,rad$^{0.77}$ and 
$B_{gb}$=(80$\pm$79)\,Mpc$^{1.77}$.
After correcting for the different cosmologies, these values 
are smaller but consistent with the clustering result of 
\citet{Wurtz1997}.
This suggests that \pks2155\ reside in a small cluster/group of
galaxy with an Abell richness class $<$0.

\subsection{Physical properties of the overdensity}\label{sec:overspec}

\begin{figure}
\centering
\includegraphics[width=0.99\columnwidth]{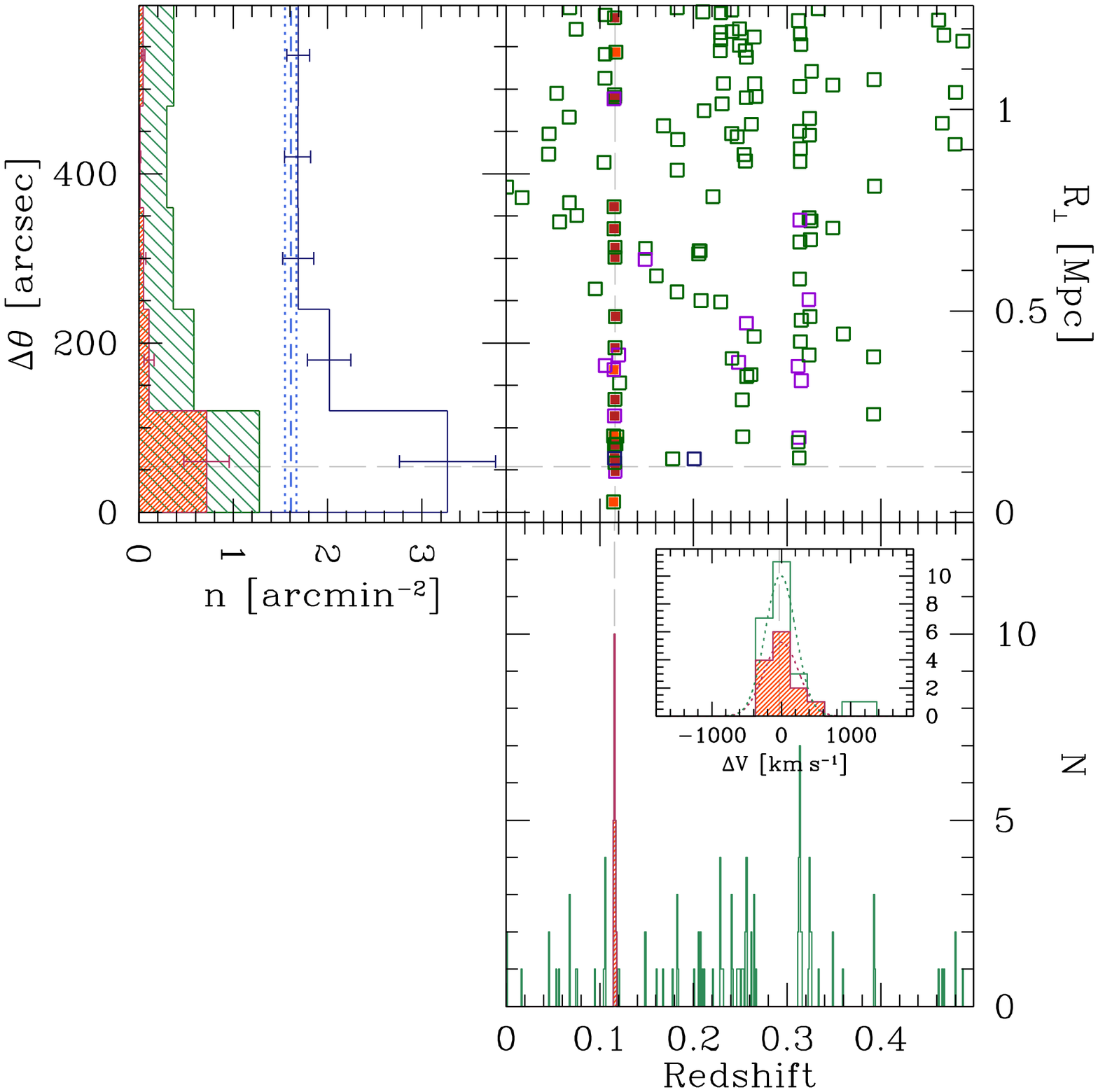}
\caption{Distribution of the sources in the
field of the \pks2155\ in the redshift -- angular 
distance from the overdensity peak plane (see Sec.\,\ref{sec:over}).
The position of \pks2155\ is marked with grey dashed
lines.  
\newline
{\it Main Panel} ---  
Green, purple, and blue squares are targets with
spectra collected by \citet{Prochaska2011}, and 
by our IMACS multi--object and long--slit spectroscopic
campaign, respectively. 
Galaxies with redshift consistent within 1000\,km\,s$^{-1}$ 
from \pks2155\ are marked with filled symbols.\newline
{\it Bottom Panel} ---
Redshift distribution of the sources (green histogram). 
The inset shows a zoom--in of galaxies within $\pm$1400\,km\,s$^{-1}$
from \pks2155.
Objects located closer than 240\arcsec\ (i.e. $\sim$500\,kpc at
z=0.116) from the peak of the overdensity are highlighted in
orange.
The best Gaussian fit of the distribution is also showed.\newline
{\it Left Panel} ---
Angular distance distribution of the sources. The blue
histogram are galaxies brighter that R$\sim$21
(i.e. M$^\star_{\rm R}$+4 at z$=$0.116). The green
shaded histogram are sources with redshift determined
spectroscopically, and the orange filled one show the 
distribution of the sources within $\pm$1000\,km\,s$^{-1}$ 
from the BL~Lac.
}\label{fig:clus}
\end{figure}

In Figure~\ref{fig:clus} we present the distribution of 
objects in the plane redshift -- angular distance from the 
luminosity centre of the group member galaxies, that
is located $\sim$70\arcsec\ South--East from
\pks2155.
% RA=21:58:54.2 and Dec.=-30:14:31).
Considering only sources brighter than M$^\star_{\rm R}$+4 
the decrease of the galaxy density at increasing angular 
distance is apparent.
The redshift distribution of the sources, in bins of $\Delta$z=0.001,
shows a clear peak at z$_{\rm g}$$=$0.11610$\pm$0.00006, consistent 
with previous estimates of the redshift of \pks2155, with a nearly 
Gaussian shape.
Assuming that only the 12~galaxies with a spectroscopic redshift
within $\pm$1000\,km\,s$^{-1}$ from z$_{\rm g}$ (not including the 
BL~Lac host galaxy) and lying within a radius of 240\arcsec\ (i.e. 
500\,kpc at z$\sim$0.116) from the luminosity centre are part of 
the overdensity, the radial component of the velocity 
dispersion is $\sigma_\parallel$=250$^{+80}_{-40}$\,km\,s$^{-1}$
\citep[where the quoted uncertainties are the 68\% confidence 
level,][]{Danese1980}. 
Under the hypothesis that the galaxy distribution traces the 
underlying mass distribution, we used the relations in 
\citet{Girardi1998} to estimated the virial radius 
(R$_{\rm vir}$$\sim$0.22\,Mpc) and the virial mass 
(M$_{\rm vir}$$\sim$1.5$\times$10$^{13}$\,M$_\odot$)
of the system.
We are cautious to take these estimates as face values.
Indeed, for small groups of galaxies, the traditional estimators 
of the virial radius and mass are not tightly correlated with their 
real values.

The presence of a second peak at z$\sim$0.314 indicates that a second 
overdensity of galaxies may be located $\sim$40\arcsec\ North from 
\pks2155.

\section{Summary and Conclusions}

We investigated the properties of the environment of \pks2155\ 
using broad band images, new multi--object spectroscopy collected 
with the IMACS instrument of the Magellan Baade telescope, and 
archival spectra gathered with the Dupont 100\arcsec\ telescope 
by \citet{Prochaska2011}.
Our measurements confirm that \pks2155\ is harboured by a 
moderate overdensity of galaxies located at z$=$0.11610$\pm$0.00006,
with a virial mass of M$_{\rm vir}$$\sim$1.5$\times$10$^{13}$\,M$_\odot$.

To our knowledge, a detailed study of the environment of BL~Lac
objects with multi object spectroscopy was performed for only 
two other targets: RGB~1745+398 \citep[z=0.267,][]{Lietzen2008}; 
and PKS~0447-439 \citep[with a possible redshift z=0.343,][]{Muriel2015}.
Both the sources are found to be associated with large galaxy
overdensities with virial masses of few times 10$^{14}$\,M$_\odot$.
In addition, the detection with Chandra of a diffuse X--ray emission, 
on scales of $\gsim$100\,kpc, around PKS~0548--322 and PKS~2005--489
suggests that these two BL~Lacs are embedded comparably massive 
galaxy clusters \citep[][]{Donato2003}.
On the contrary, the potentially presence of gravitational arcs 
in the HST images of H~1517+656 \citep[z=0.702\footnote{The 
absorption lines used to determine the redshift could belong 
to an intervening system, rather than to the BL~Lac host,
so z=0.702 should be considered as a firm lower limit for 
the redshift of the source \citep[see Sec.\,3 in][]{Beckmann1999}.},][]{Scarpa1999} 
allowed \citet{Beckmann1999} to estimate a 
virial mass more similar to what we observed for \pks2155: 
M$_{\rm vir}$$\sim$1.4$\times$10$^{13}$\,M$_\odot$.
This (small) sample for which the virial mass could be estimated, 
indicates that BL~Lac objects are typically hosted by haloes 
spanning a range of masses, from $\sim$10$^{13}$\,M$_\odot$ 
to few times 10$^{14}$\,M$_\odot$.
This suggests that radio--loud sources, in particular BL Lac 
objects, may be found in a variety (although still massive) 
galactic environments \citep[see also][]{Wurtz1997, Lietzen2011}.
Although based on a small and heterogeneous sample, it is 
interesting to put this result in the contest of the current 
merger driven paradigm of the nuclear activity \citep[e.g.][]{Barnes1991, 
Dimatteo2005}.
In small groups, major mergers between gas rich galaxies are 
expected to be frequent and particularly effective in funnelling
huge amount of gas in the nuclear regions, eventually triggering 
the central black holes activity \citep[e.g.][]{Hopkins2008}.
Conversely, in rich clusters, the high--velocity dispersion 
decreases the effective cross section of galaxy interactions, 
in spite of the higher galaxy densities \citep[e.g.][]{Aarseth1980}.
Galaxy harassment due to high--speed interactions (but not
merger) of galaxies could drive dynamical instabilities that
efficiently channel gas onto the super--massive black holes
in rich environment \citep[][]{Moore1996}. 
This requires the host galaxy to contains huge gas reservoir.
However, ram--pressure stripping is likely to remove the
cold gas in cluster galaxies, and the BL~Lac hosts are 
found to be deficient in molecular gas when compared with
quasars \citep[e.g.][]{Fumagalli2012}.
The trigger mechanism(s) for the nuclear activity of the
BL~Lac host galaxies is thus likely to be complex and not 
universal.
A larger sample of BL~Lac with a detailed characterisation
of the large scale environment is thus needed to understand
this process in a statistical manner.

% This confirms that the radio loud luminosity of AGN is 
% not related with the environmental properties of the 
% host galaxies, with the brightest radio emitters (i.e.
% the BL~Lacs) that are hosted by haloes with masses
% similar to those of less luminous radio--loud AGN 
% \citep[e.g.][]{Mandelbaum2009, Worpel2013}.

% This is in agreement with recent numerical simulations that 
% predict the most powerfull radio--loud AGN to be hosted by 
% high mass haloes (e.g., Fanidakis et al., in prep.). 

% 1517+656 \citep[][]{Beckmann1999};

\section*{Acknowledgements}

EPF acknowledges funding through the ERC grant `Cosmic Dawn'.
MF acknowledges support by the Science and Technology Facilities 
Council, grant number ST/L00075X/1.
This paper includes data gathered with the 6.5\,meter Magellan 
Telescopes located at Las Campanas Observatory (Chile). We 
thank J.~O'Meara for assisting during observations. EPF 
is grateful to T.~Schmidt, E.~Ba{\~n}ados, A.~Obreja, M.~Fouesneau, 
and C.~Ferkinhoff for providing support in the use of \texttt{python} 
for the data reduction. 
The authors thank A.~Treves and R.~Falomo for their important 
contributions to data interpretation and to the shaping of the 
article.
The authors acknowledge C.~Mazzucchelli and M.~Landoni for useful 
comments and suggestions.
This research made use of \texttt{Astropy}, a community--developed 
core Python package for Astronomy \citep{Astropy2013} and of
\texttt{APLpy}, an open-source plotting package for \texttt{python}\footnote{\texttt{http://aplpy.github.io/}}.

\section*{Supporting Information}

Spectra of the sources collected through multy--object and long--slit 
spectroscopy are available in the online version of this article.

\begin{figure*}
\centering
\includegraphics[width=1.99\columnwidth]{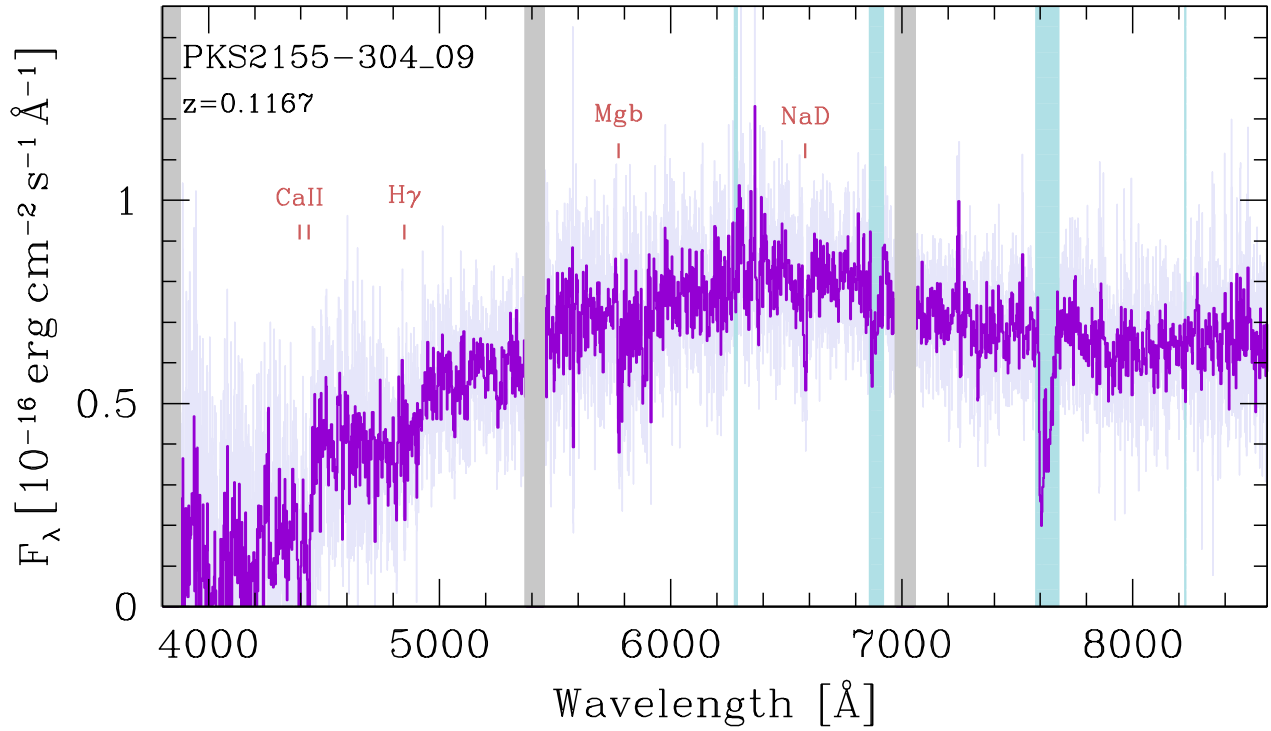}\newline
\vspace{.5cm}\\
\includegraphics[width=1.99\columnwidth]{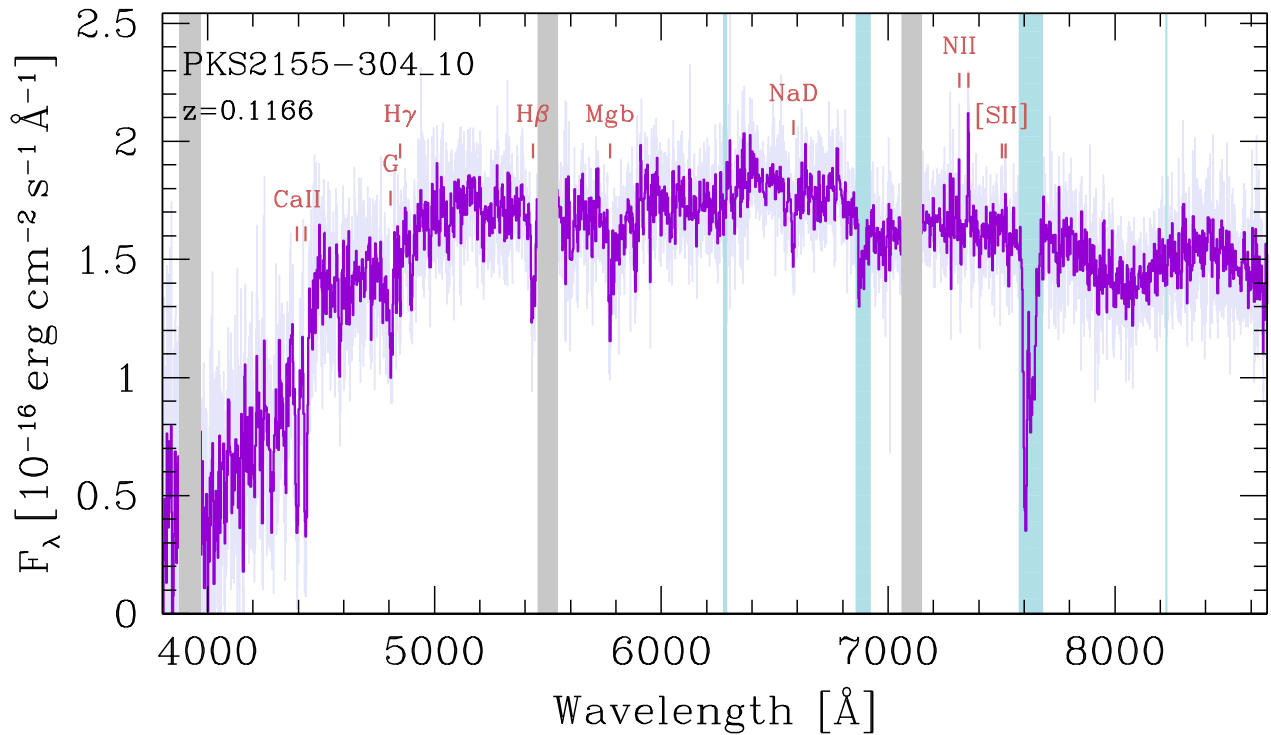}\newline
\vspace{.5cm}\\
\caption{
Spectra of the sources collected through multy--object spectroscopy
corrected for Galactic extinction (pale purple line) and binned by 
2\AA\ (dark purple). Spectral features used to determine the 
redshift of the sources are labelled. Shaded grey regions mark the 
gaps among the IMACS's ccds, while the light blue ones show the 
position of the most prominent telluric absorptions.}\label{fig:mos}
\end{figure*}

\addtocounter{figure}{-1}

\begin{figure*}
\centering
\includegraphics[width=1.99\columnwidth]{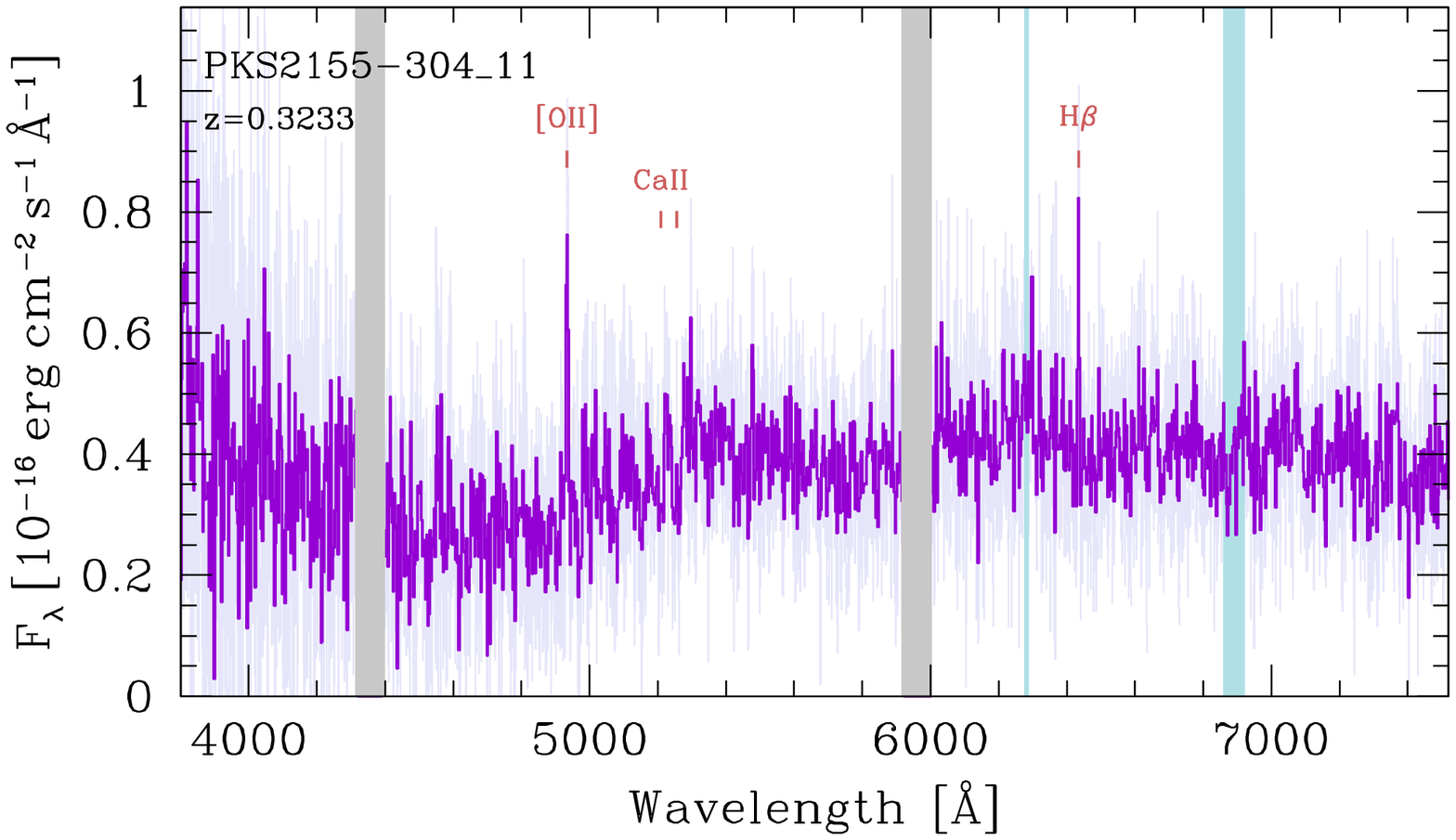}\newline
\vspace{.5cm}\\
\includegraphics[width=1.99\columnwidth]{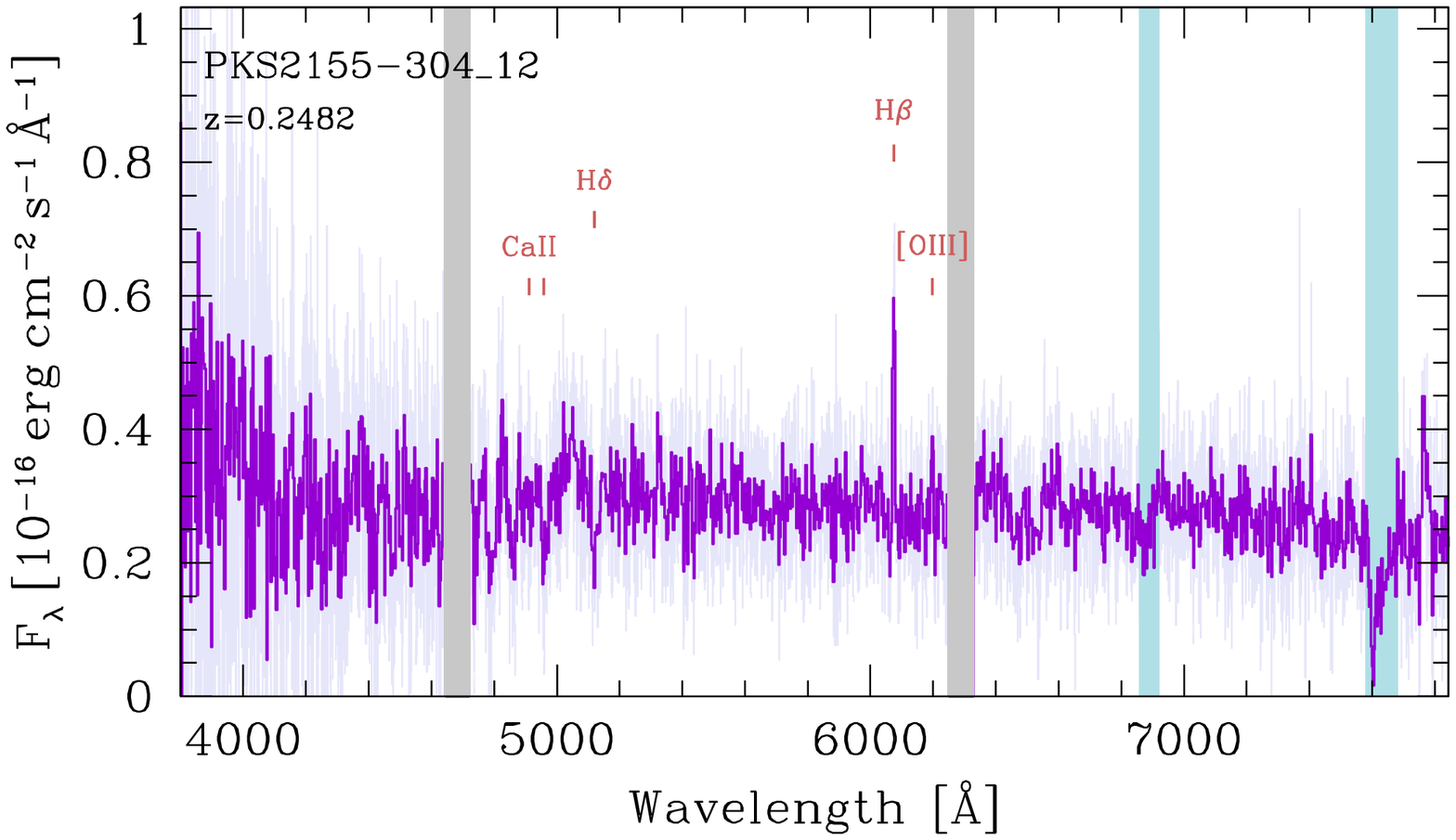}\newline
\vspace{.5cm}\\
\caption{ continued.}
\end{figure*}

\addtocounter{figure}{-1}

\begin{figure*}
\centering
\includegraphics[width=1.99\columnwidth]{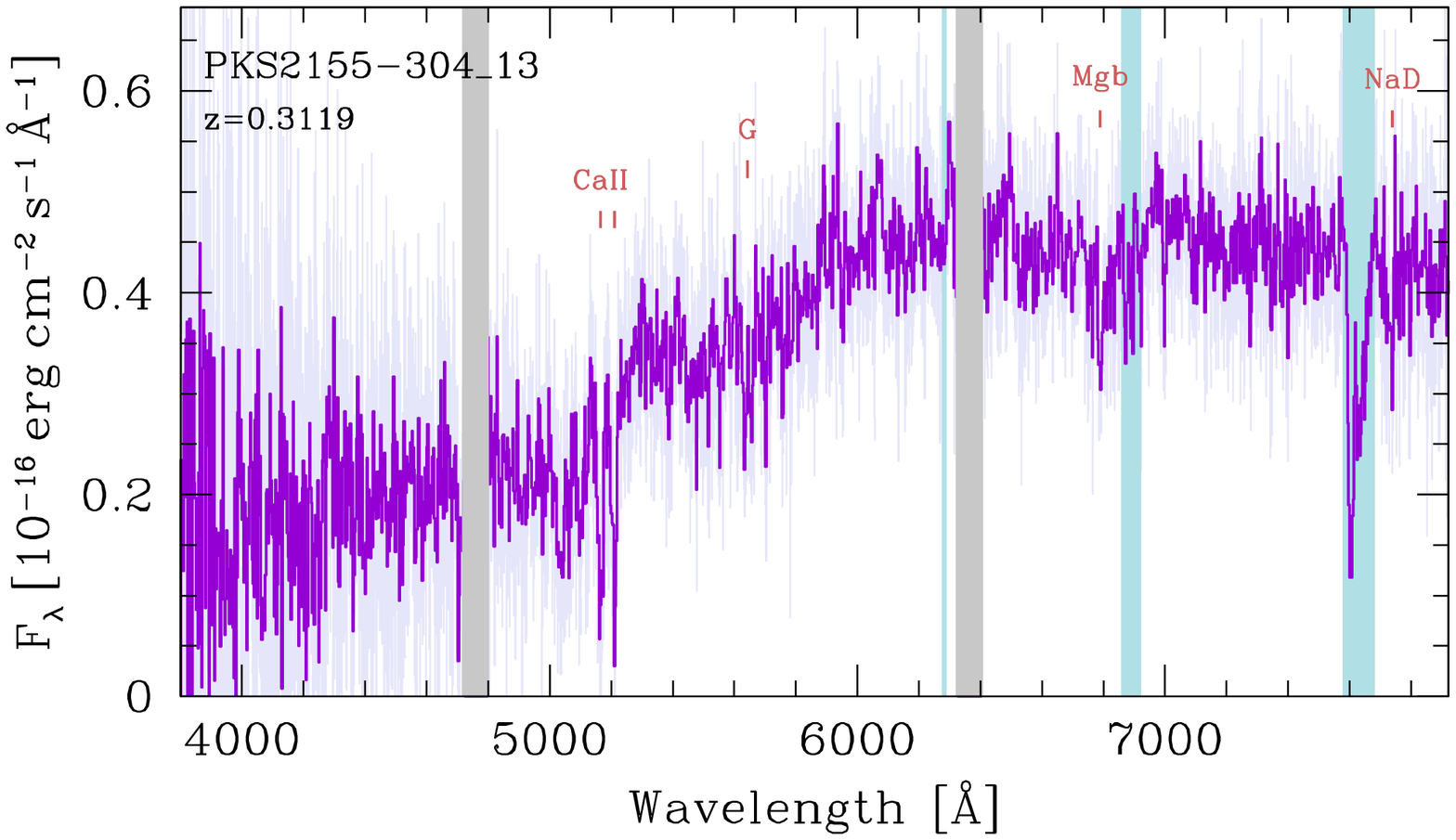}\newline
\vspace{.5cm}\\
\includegraphics[width=1.99\columnwidth]{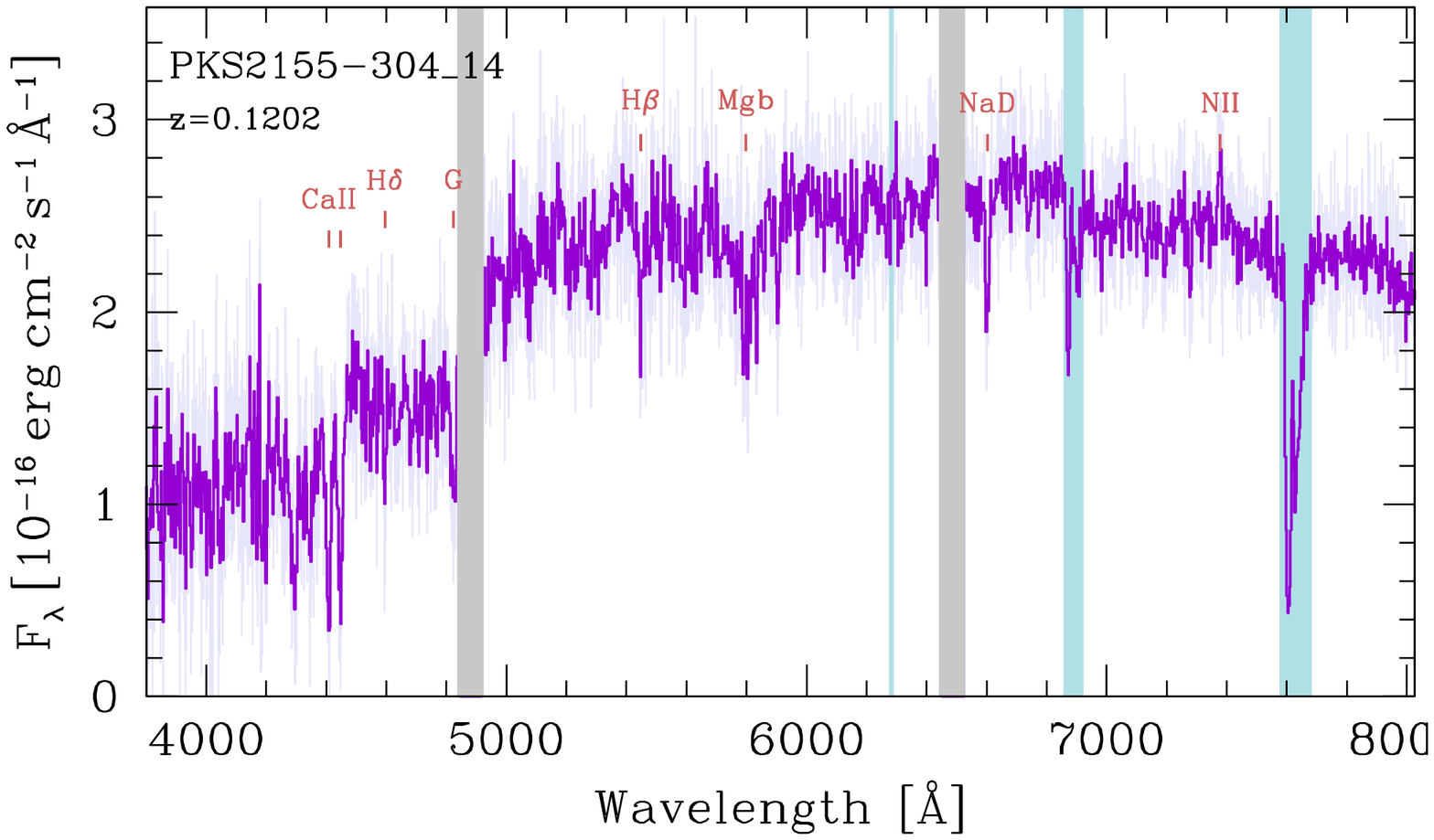}\newline
\vspace{.5cm}\\
\caption{ continued.}
\end{figure*}

\addtocounter{figure}{-1}

\begin{figure*}
\centering
\includegraphics[width=1.99\columnwidth]{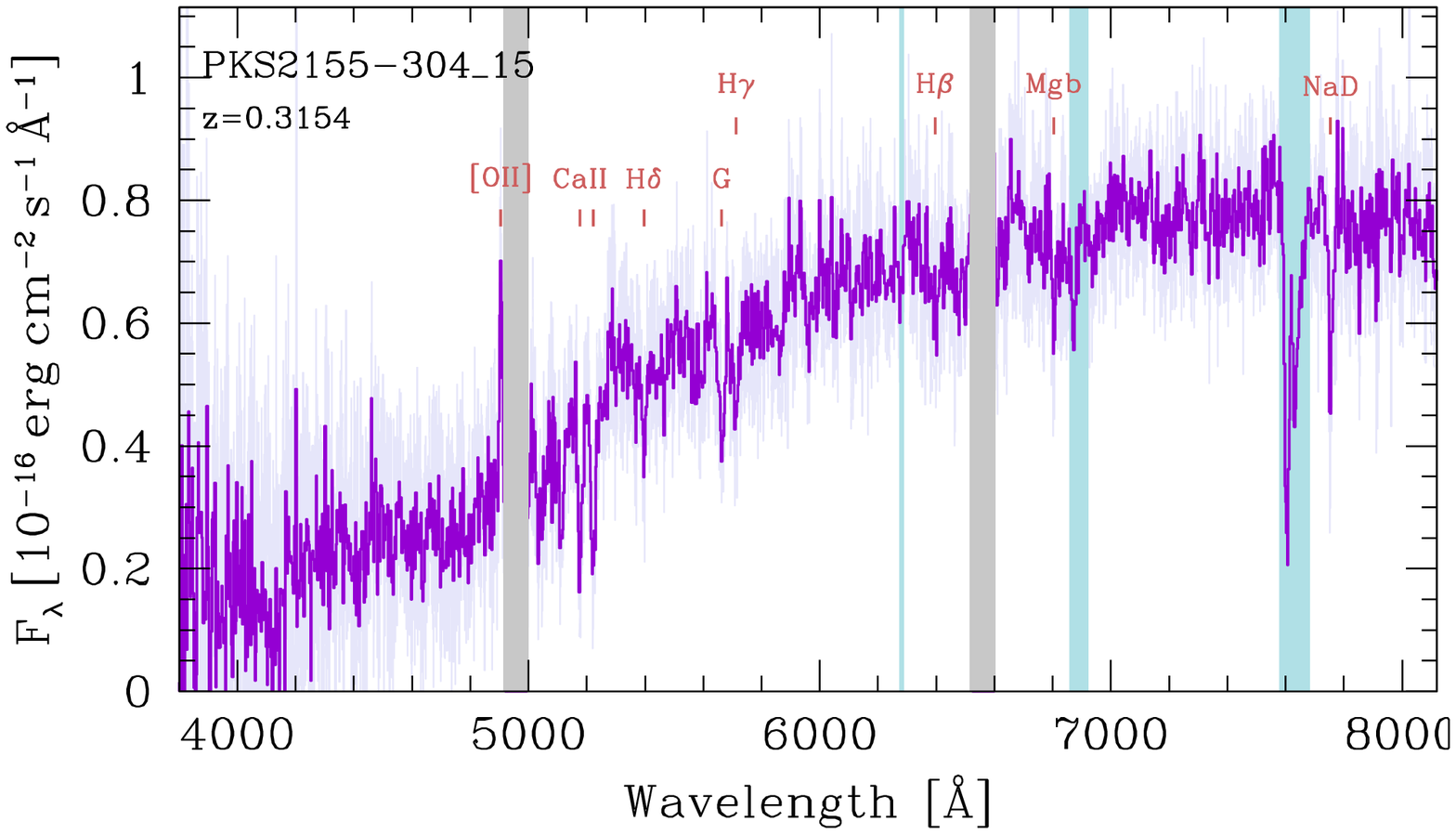}\newline
\vspace{.5cm}\\
\includegraphics[width=1.99\columnwidth]{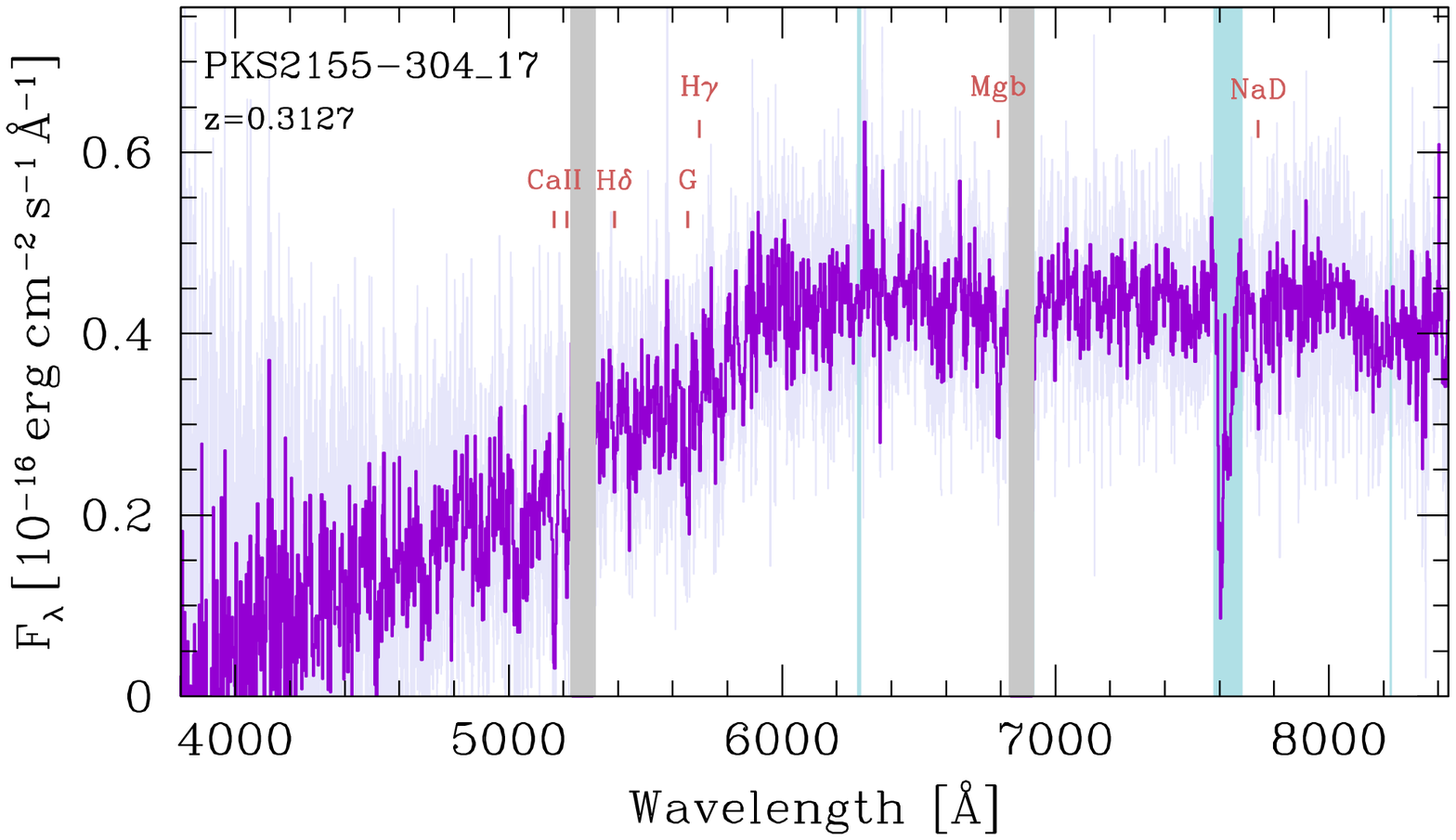}\newline
\vspace{.5cm}\\
\caption{ continued.}
\end{figure*}

\addtocounter{figure}{-1}

\begin{figure*}
\centering
\includegraphics[width=1.99\columnwidth]{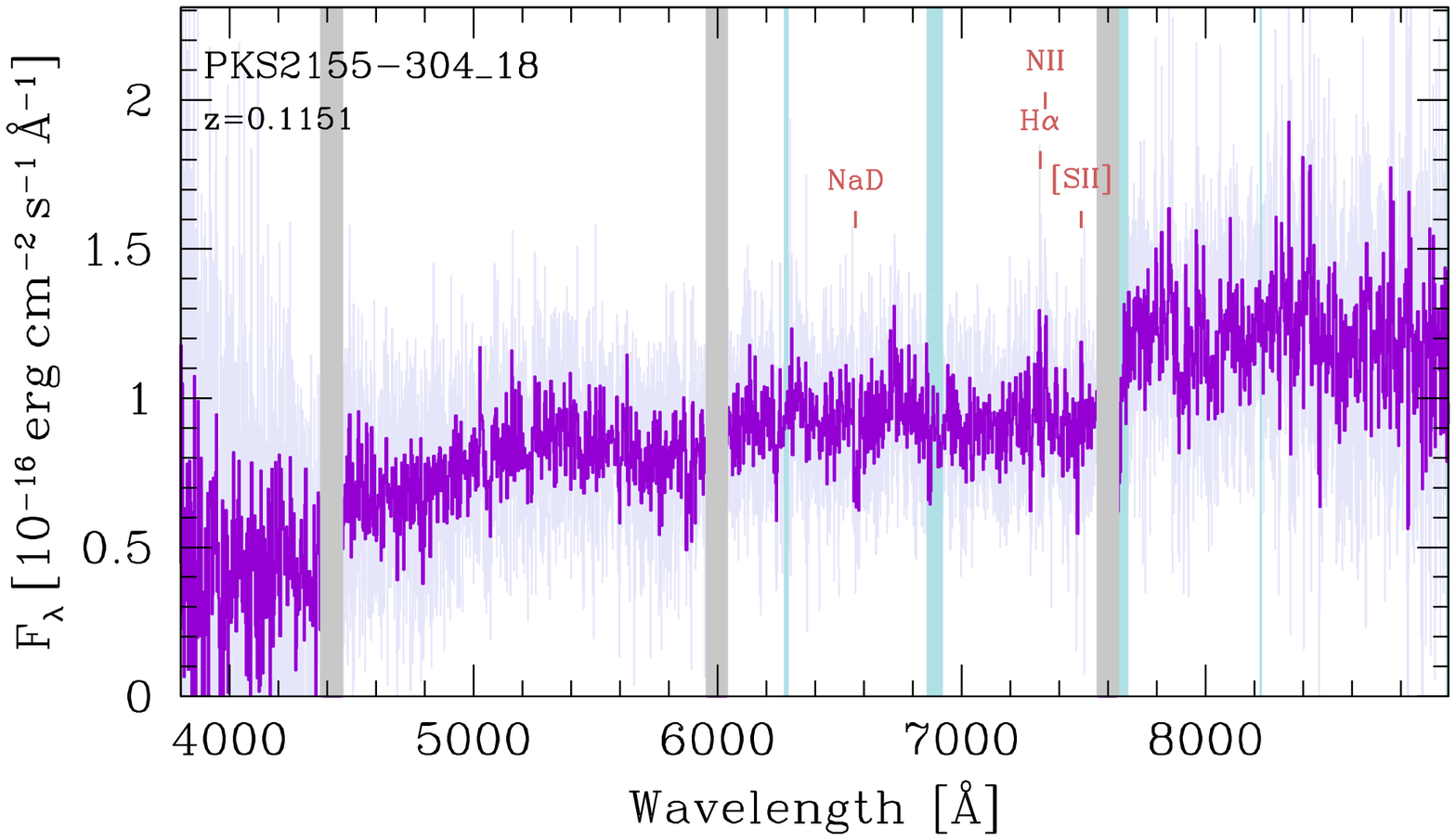}\newline
\vspace{.5cm}\\
\includegraphics[width=1.99\columnwidth]{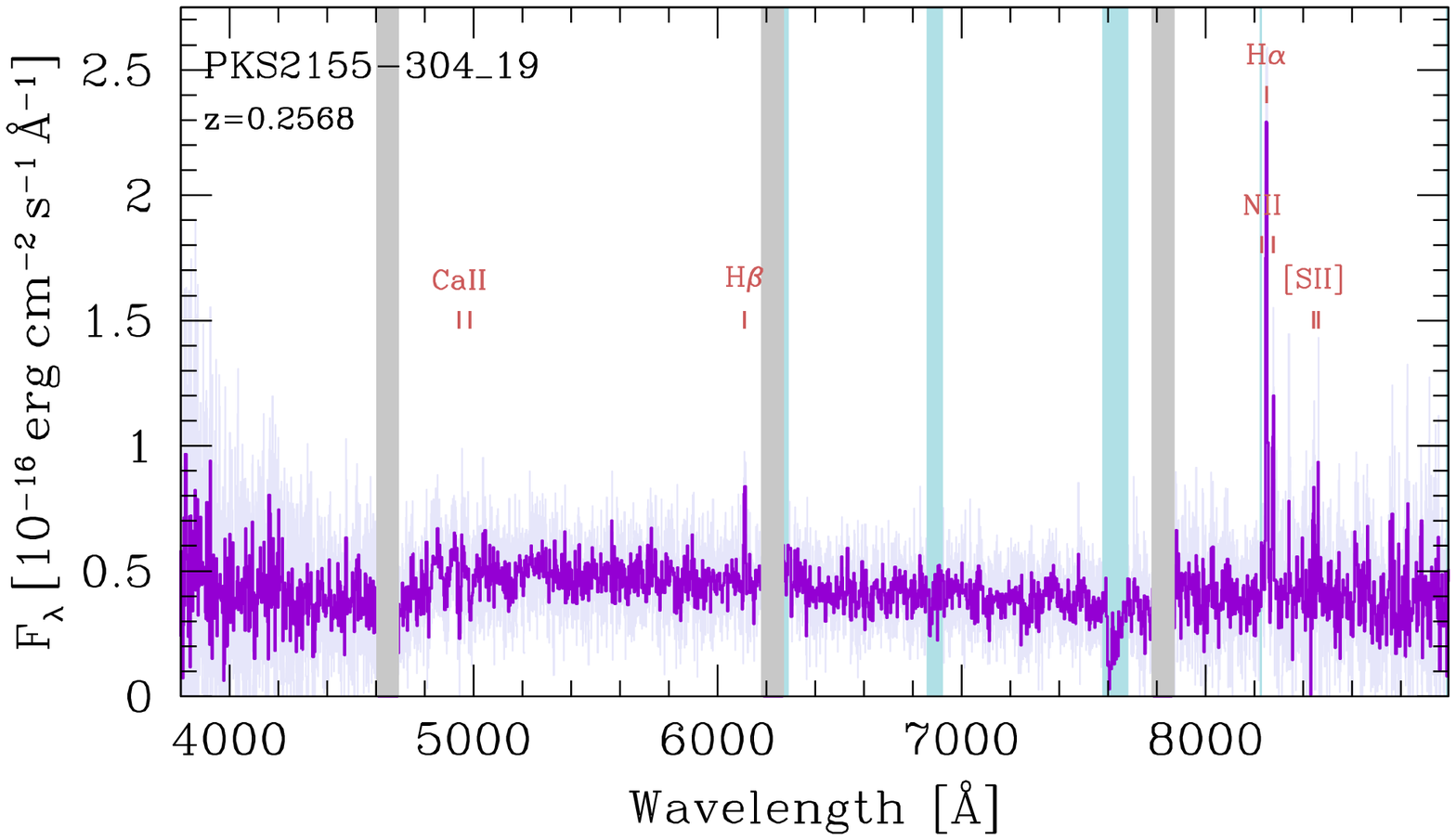}\newline
\vspace{.5cm}\\
\caption{ continued.}
\end{figure*}

\addtocounter{figure}{-1}

\begin{figure*}
\centering
\includegraphics[width=1.99\columnwidth]{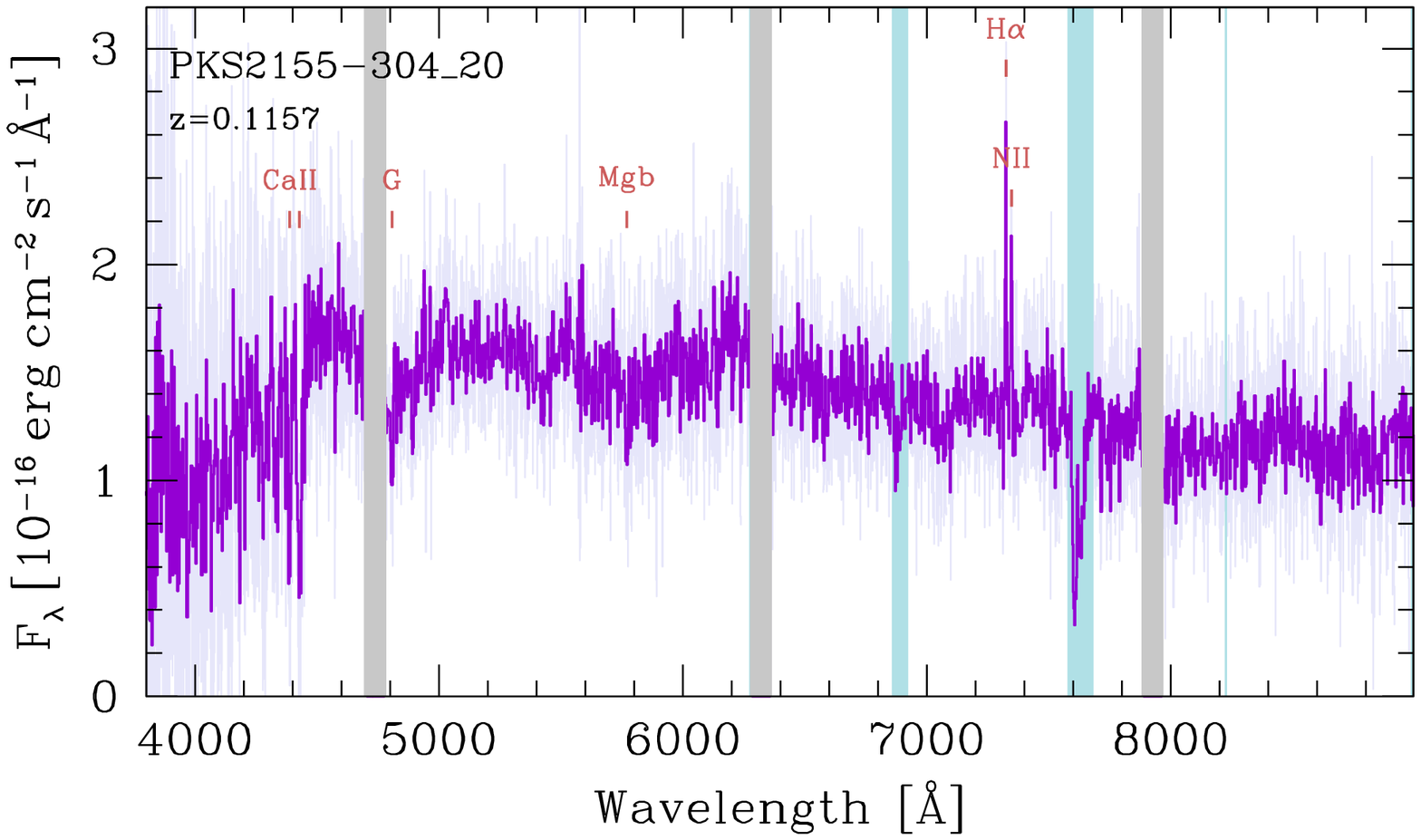}\newline
\vspace{.5cm}\\
\includegraphics[width=1.99\columnwidth]{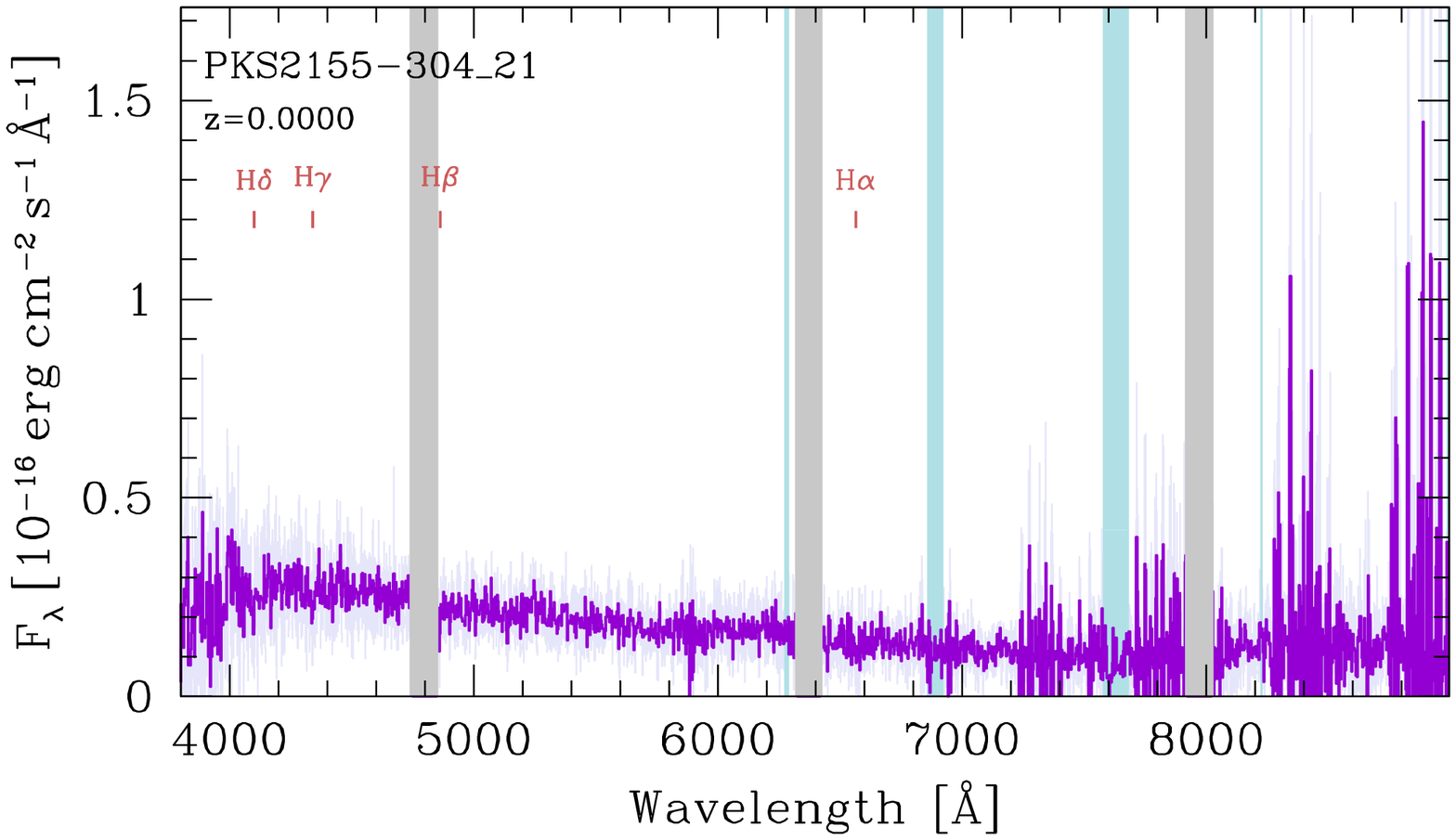}\newline
\vspace{.5cm}\\
\caption{ continued.}
\end{figure*}

\addtocounter{figure}{-1}

\begin{figure*}
\centering
\includegraphics[width=1.99\columnwidth]{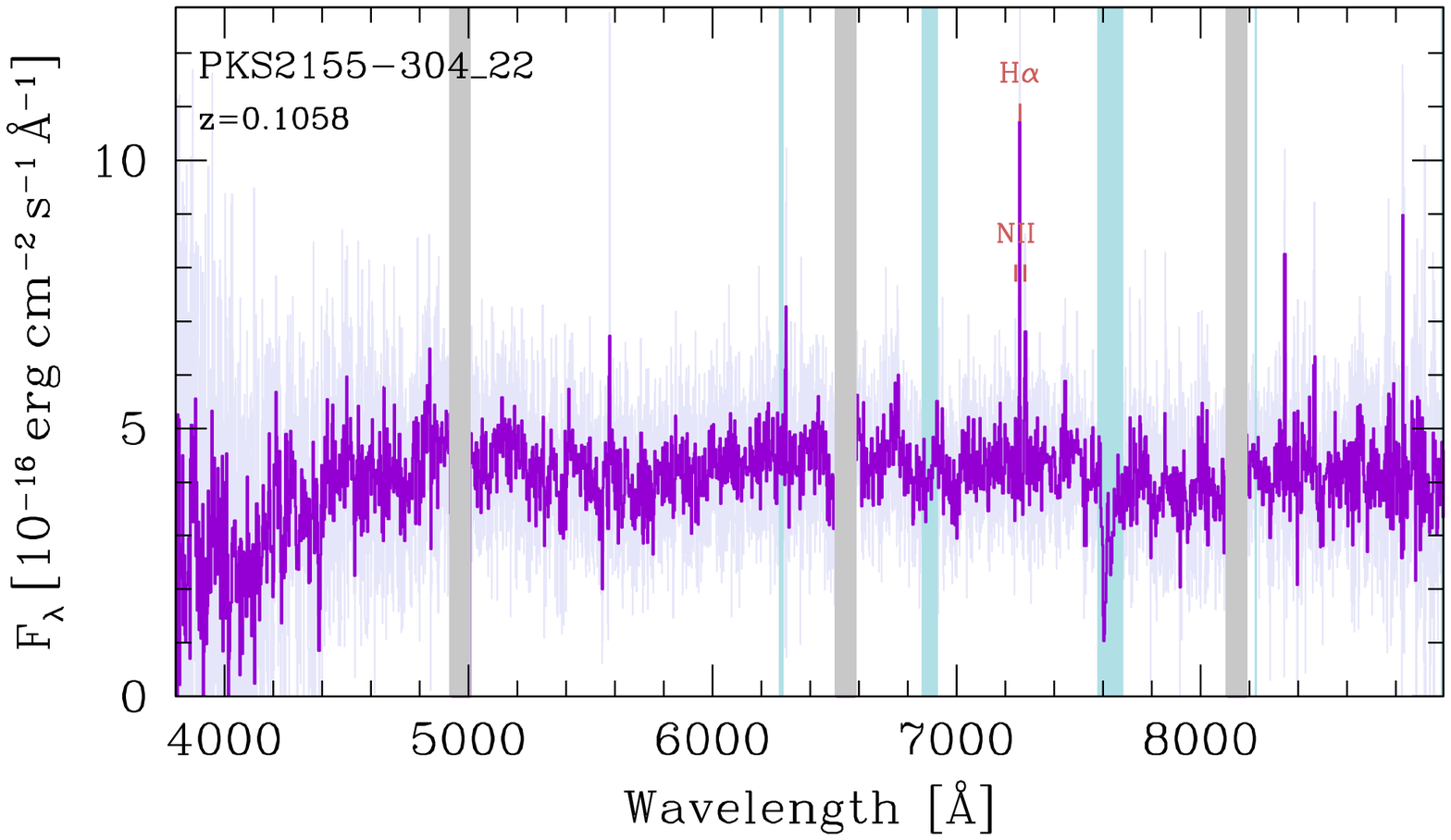}\newline
\vspace{.5cm}\\
\includegraphics[width=1.99\columnwidth]{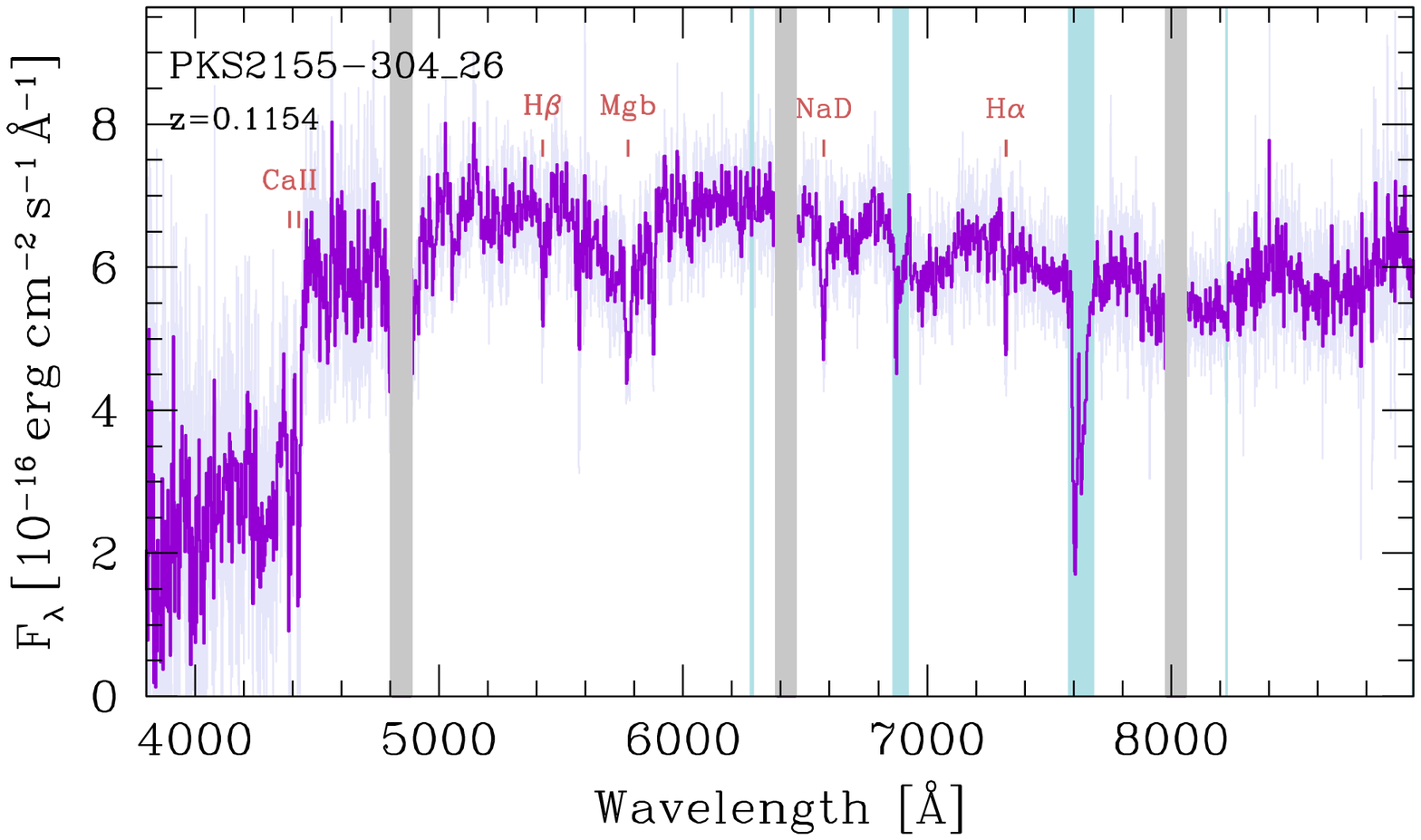}\newline
\vspace{.5cm}\\
\caption{ continued.}
\end{figure*}

\addtocounter{figure}{-1}

\begin{figure*}
\centering
\includegraphics[width=1.99\columnwidth]{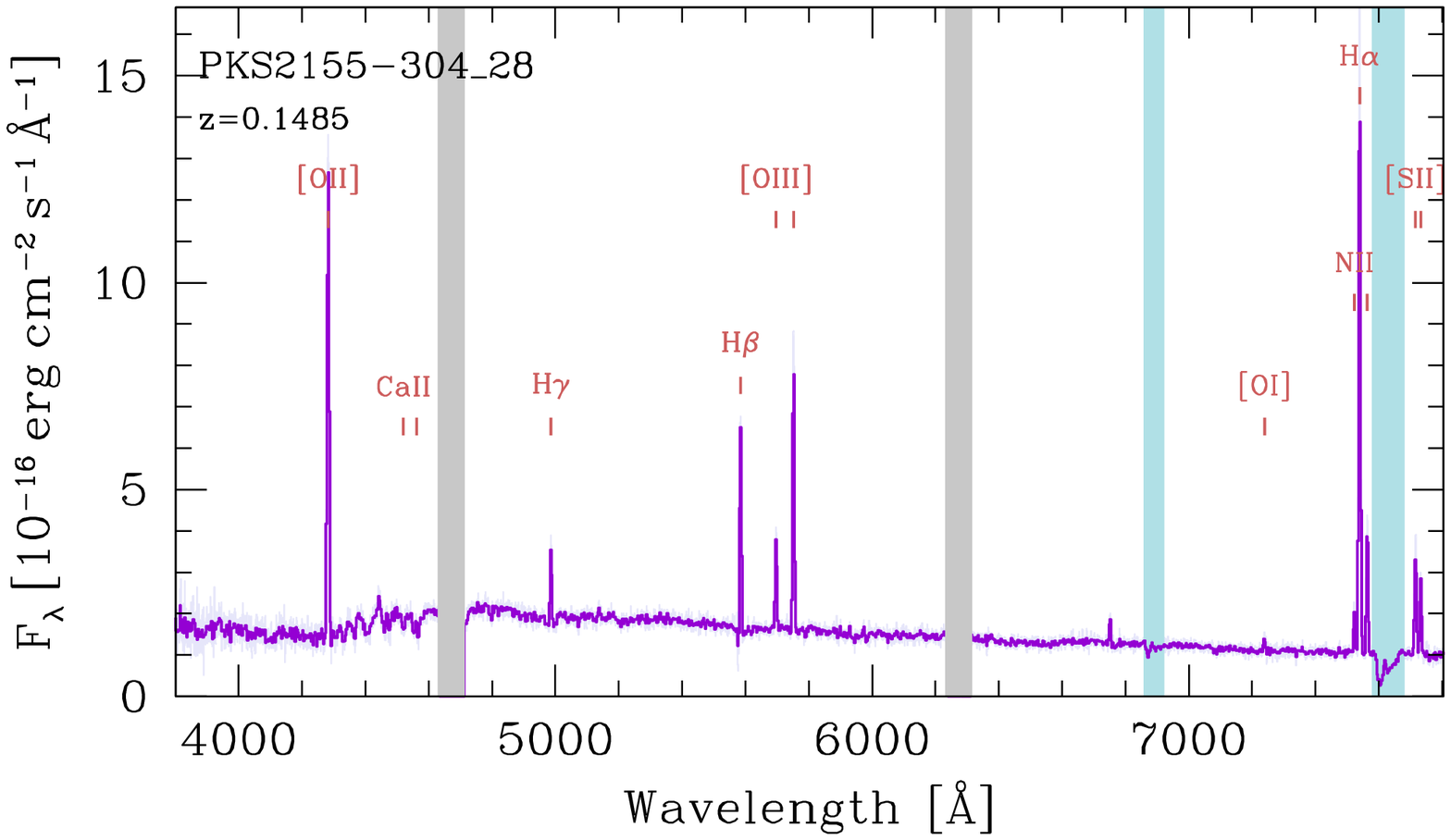}\newline
\vspace{.5cm}\\
\includegraphics[width=1.99\columnwidth]{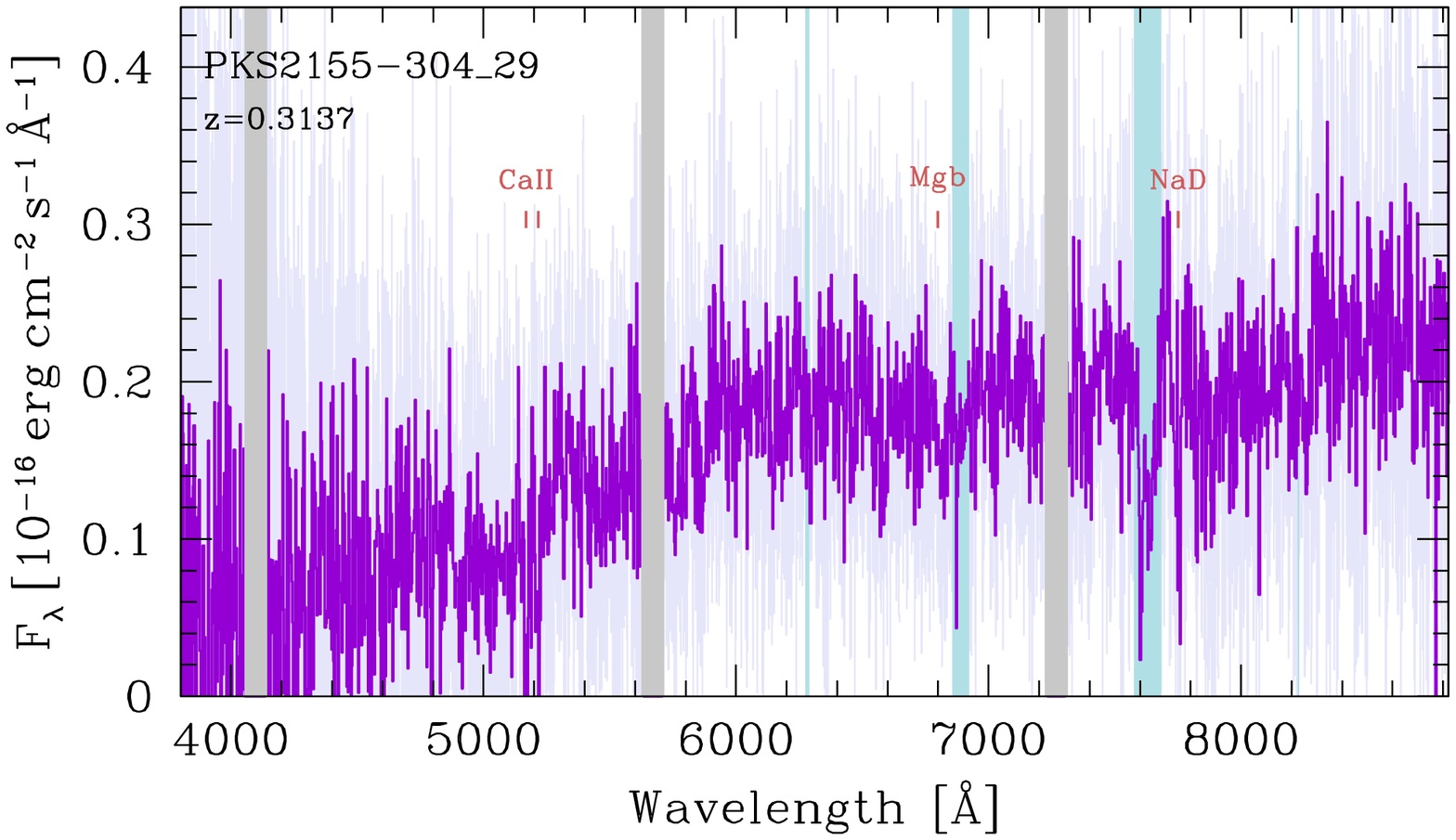}\newline
\vspace{.5cm}\\
\caption{ continued.}
\end{figure*}

\begin{figure*}
\centering
\includegraphics[width=1.99\columnwidth]{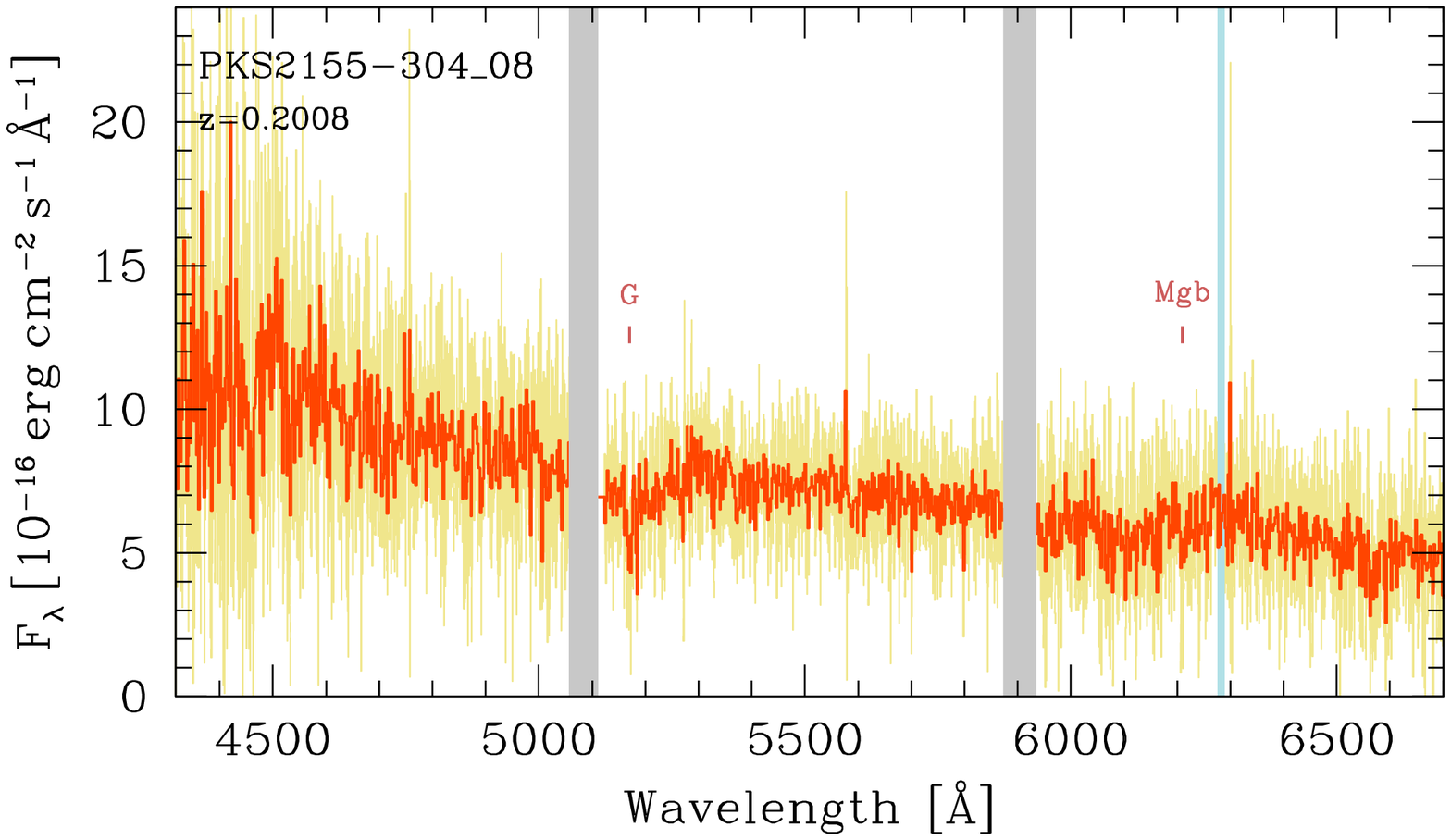}\newline
\vspace{.5cm}\\
\includegraphics[width=1.99\columnwidth]{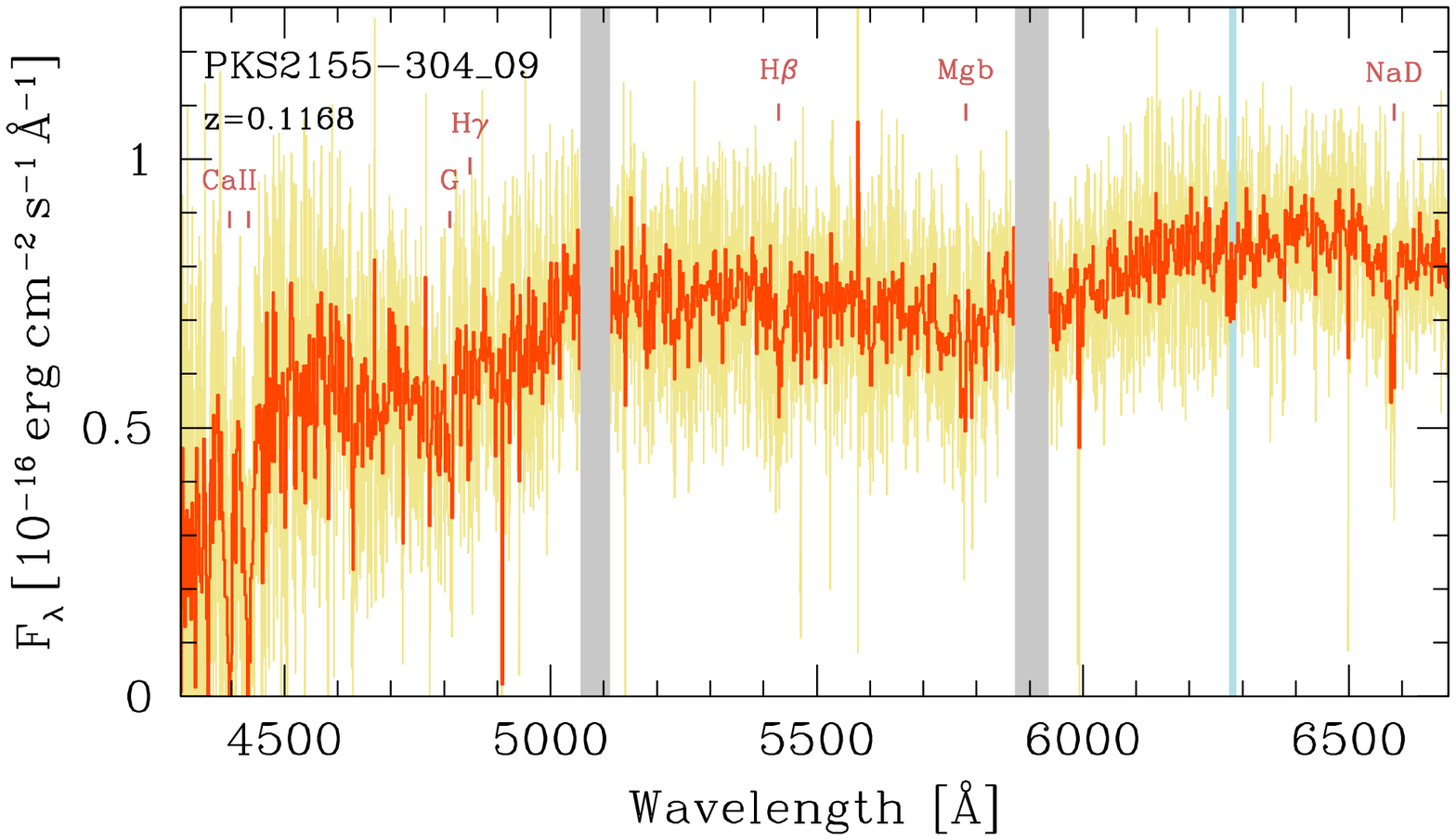}\newline
\vspace{.5cm}\\
\caption{Spectra of the sources collected through 
long--slit spectroscopy corrected for Galactic extinction (yellow line) 
and binned by 2\AA\ (dark orange). Labels and shaded regions are 
equivalent to those of Figure~\ref{fig:mos}.}\label{fig:lon}
\end{figure*}

\begin{figure*}
\centering
\includegraphics[width=1.99\columnwidth]{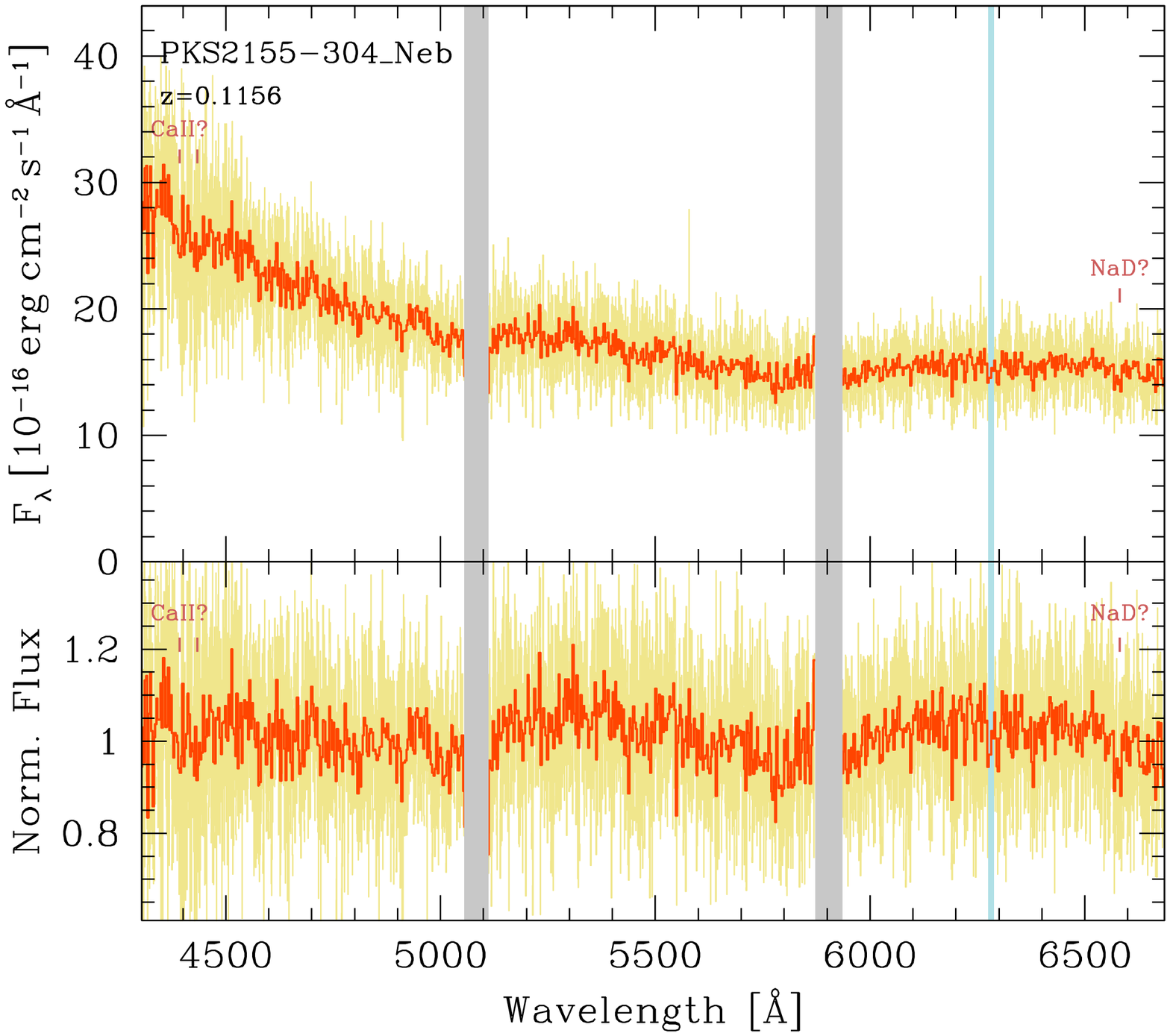}
\caption{
Spectra of the sources collected through 
long--slit spectroscopy corrected for Galactic extinction (yellow line) 
and binned by 2\AA\ (dark orange). Labels and shaded regions are 
equivalent to those of Figure~\ref{fig:mos}.}\label{fig:neb}
\end{figure*}

\label{lastpage}


\begin{thebibliography}{99}


\bibitem[\protect\citeauthoryear{Aarseth \& Fall}{1980}]{Aarseth1980} Aarseth S.~J., Fall S.~M., 1980, ApJ, 236, 43 



\bibitem[\protect\citeauthoryear{Aharonian et al.}{2005}]{Aharonian2005} Aharonian F., et al., 2005, A\&A, 430, 865 
\bibitem[\protect\citeauthoryear{Aharonian et al.}{2007}]{Aharonian2007} Aharonian F., et al., 2007, ApJ, 664, L71
\bibitem[\protect\citeauthoryear{Aharonian et al.}{2013}]{Aharonian2013} Aharonian F., Essey W., Kusenko A., Prosekin A., 2013, PhRvD, 87, 063002 
\bibitem[\protect\citeauthoryear{Astropy Collaboration et al.}{2013}]{Astropy2013} Astropy Collaboration, et al., 2013, A\&A, 558, A33 
\bibitem[\protect\citeauthoryear{Bahcall}{1981}]{Bahcall1981} Bahcall N.~A., 1981, ApJ, 247, 787 

\bibitem[\protect\citeauthoryear{Barnes \& Hernquist}{1991}]{Barnes1991} Barnes J.~E., Hernquist L.~E., 1991, ApJ, 370, L65 
\bibitem[\protect\citeauthoryear{Beckmann, Bade, \& Wucknitz}{1999}]{Beckmann1999} Beckmann V., Bade N., Wucknitz O., 1999, A\&A, 352, 395 
\bibitem[\protect\citeauthoryear{Bertin \& Arnouts}{1996}]{Bertin1996} Bertin E., Arnouts S., 1996, A\&AS, 117, 393 
\bibitem[\protect\citeauthoryear{Blades, Hunstead, \& Murdoch}{1981}]{Blades1981} Blades J.~C., Hunstead R.~W., Murdoch H.~S., 1981, MNRAS, 194, 669 
\bibitem[\protect\citeauthoryear{Blandford \& Rees}{1978}]{Blandford1978} Blandford R.~D., Rees M.~J., 1978, bllo.conf, 328
\bibitem[\protect\citeauthoryear{Blanton et al.}{2001}]{Blanton2001} Blanton M.~R., et al., 2001, AJ, 121, 2358 
\bibitem[\protect\citeauthoryear{Boksenberg \& Sargent}{1978}]{Boksenberg1978} Boksenberg A., Sargent W.~L.~W., 1978, ApJ, 220, 42 


\bibitem[\protect\citeauthoryear{Chadwick et al.}{1999a}]{Chadwick1999a} Chadwick P.~M., et al., 1999, APh, 11, 145 
\bibitem[\protect\citeauthoryear{Chadwick et al.}{1999b}]{Chadwick1999b} Chadwick P.~M., et al., 1999, ApJ, 513, 
161 
\bibitem[\protect\citeauthoryear{Decarli et al.}{2008}]{Decarli2008} Decarli R., Labita M., Treves A., Falomo R., 2008, MNRAS, 387, 1237 
\bibitem[\protect\citeauthoryear{Danese, de Zotti, \& di Tullio}{1980}]{Danese1980} Danese L., de Zotti G., di Tullio G., 1980, A\&A, 82, 322
\bibitem[\protect\citeauthoryear{Dekker, Delabre, \& Dodorico}{1986}]{Dekker1986} Dekker H., Delabre B., Dodorico S., 1986, SPIE, 627, 339 

\bibitem[\protect\citeauthoryear{Di Matteo, Springel, \& Hernquist}{2005}]{Dimatteo2005} Di Matteo T., Springel V., Hernquist L., 2005, Natur, 433, 604 


\bibitem[\protect\citeauthoryear{Donato et al.}{2003}]{Donato2003} Donato D., Gliozzi M., Sambruna R.~M., Pesce J.~E., 2003, A\&A, 407, 503 
\bibitem[\protect\citeauthoryear{Dressler et al.}{2011}]{Dressler2011} Dressler A., et al., 2011, PASP, 123, 288 
\bibitem[\protect\citeauthoryear{Falomo, Melnick, \& Tanzi}{1990}]{Falomo1990} Falomo R., Melnick J., Tanzi E.~G., 1990, Natur, 345, 692 
\bibitem[\protect\citeauthoryear{Falomo et al.}{1991}]{Falomo1991} Falomo R., Giraud E., Melnick J., Maraschi L., Tanzi E.~G., Treves A., 1991, ApJ, 380, L67 
\bibitem[\protect\citeauthoryear{Falomo, Pesce, \& Treves}{1993}]{Falomo1993} Falomo R., Pesce J.~E., Treves A., 1993, ApJ, 411, L63 
\bibitem[\protect\citeauthoryear{Falomo, Pesce, \& Treves}{1995}]{Falomo1995} Falomo R., Pesce J.~E., Treves A., 1995, ApJ, 438, L9 \bibitem[\protect\citeauthoryear{Falomo}{1996}]{Falomo1996} Falomo R., 1996, MNRAS, 283, 241 
\bibitem[\protect\citeauthoryear{Falomo \& Kotilainen}{1999}]{Falomo1999} Falomo R., Kotilainen J.~K., 1999, A\&A, 352, 85
\bibitem[\protect\citeauthoryear{Falomo, Pian, \& Treves}{2014}]{Falomo2014} Falomo R., Pian E., Treves A., 2014, A\&ARv, 22, 73 
\bibitem[\protect\citeauthoryear{Fan \& Lin}{2000}]{Fan2000} Fan J.~H., Lin R.~G., 2000, A\&A, 355, 880 
\bibitem[\protect\citeauthoryear{Fumagalli et al.}{2012}]{Fumagalli2012} Fumagalli M., Dessauges-Zavadsky M., Furniss A., Prochaska J.~X., Williams D.~A., Kaplan K., Hogan M., 2012, MNRAS, 424, 2276 

\bibitem[\protect\citeauthoryear{Gavazzi et al.}{2003}]{Gavazzi2003} Gavazzi G., Boselli A., Donati A., Franzetti P., Scodeggio M., 2003, A\&A, 400, 451 
\bibitem[\protect\citeauthoryear{Girardi et al.}{1998}]{Girardi1998} Girardi M., Giuricin G., Mardirossian F., Mezzetti M., Boschin W., 1998, ApJ, 505, 74 
\bibitem[\protect\citeauthoryear{Gladders \& Yee}{2000}]{Gladders2000} Gladders M.~D., Yee H.~K.~C., 2000, AJ, 120, 2148

\bibitem[\protect\citeauthoryear{Griffiths et al.}{1979}]{Griffiths1979} Griffiths R.~E., Briel U., Chaisson L., Tapia S., 1979, ApJ, 234, 810 

\bibitem[\protect\citeauthoryear{Hopkins et al.}{2008}]{Hopkins2008} Hopkins P.~F., Hernquist L., Cox T.~J., Kere{\v s} D., 2008, ApJS, 175, 356 

\bibitem[\protect\citeauthoryear{Kelson}{2003}]{Kelson2003} Kelson D.~D., 2003, PASP, 115, 688 
\bibitem[\protect\citeauthoryear{Kotilainen, Falomo, \& Scarpa}{1998}]{Kotilainen1998} Kotilainen J.~K., Falomo R., Scarpa R., 1998, A\&A, 336, 479 

\bibitem[\protect\citeauthoryear{Kotilainen et al.}{2011}]{Kotilainen2011} Kotilainen J.~K., Hyv{\"o}nen T., Falomo R., Treves A., Uslenghi M., 2011, A\&A, 534, L2 
\bibitem[\protect\citeauthoryear{Kinney et al.}{1996}]{Kinney1996} Kinney A.~L., Calzetti D., Bohlin R.~C., McQuade K., Storchi-Bergmann T., Schmitt H.~R., 1996, ApJ, 467, 38 
\bibitem[\protect\citeauthoryear{Lang et al.}{2010}]{Lang2010} Lang D., Hogg D.~W., Mierle K., Blanton M., Roweis S., 2010, AJ, 139, 1782
\bibitem[\protect\citeauthoryear{Lietzen et al.}{2008}]{Lietzen2008} Lietzen H., Nilsson K., Takalo L.~O., Hein{\"a}m{\"a}ki P., Nurmi P., Kein{\"a}nen P., Wagner S., 2008, A\&A, 482, 771 
\bibitem[\protect\citeauthoryear{Lietzen et al.}{2011}]{Lietzen2011} Lietzen H., Hein{\"a}m{\"a}ki P., Nurmi P., Liivam{\"a}gi L.~J., Saar E., Tago E., Takalo L.~O., Einasto M., 2011, A\&A, 535, A21 


\bibitem[\protect\citeauthoryear{Liuzzo et al.}{2013}]{Liuzzo2013} Liuzzo E., et al., 2013, AJ, 145, 73 

\bibitem[\protect\citeauthoryear{Longair \& Seldner}{1979}]{Longair1979} Longair M.~S., Seldner M., 1979, MNRAS, 189, 433 

\bibitem[\protect\citeauthoryear{L{\'o}pez-Cruz, Barkhouse, \& Yee}{2004}]{Lopez2004} L{\'o}pez-Cruz O., Barkhouse W.~A., Yee H.~K.~C., 2004, ApJ, 614, 679 
\bibitem[\protect\citeauthoryear{Mandelbaum et al.}{2009}]{Mandelbaum2009} Mandelbaum R., Li C., Kauffmann G., White S.~D.~M., 2009, MNRAS, 393, 377 
\bibitem[\protect\citeauthoryear{Moore et al.}{1996}]{Moore1996} Moore B., Katz N., Lake G., Dressler A., Oemler A., 1996, Natur, 379, 613 
\bibitem[\protect\citeauthoryear{Muriel et al.}{2015}]{Muriel2015} Muriel H., Donzelli C., Rovero A.~C., Pichel A., 2015, A\&A, 574, A101 
\bibitem[\protect\citeauthoryear{Nilsson et al.}{2008}]{Nilsson2008} Nilsson K., Pursimo T., Sillanp{\"a}{\"a} A., Takalo L.~O., Lindfors E., 2008, A\&A, 487, L29
\bibitem[\protect\citeauthoryear{Oh et al.}{2011}]{Oh2011} Oh K., Sarzi M., Schawinski K., Yi S.~K., 2011, ApJS, 195, 13 
\bibitem[\protect\citeauthoryear{Oke}{1974}]{Oke1974} Oke J.~B., 1974, ApJS, 27, 21 
\bibitem[\protect\citeauthoryear{Oke \& Gunn}{1983}]{Oke1983} Oke J.~B., Gunn J.~E., 1983, ApJ, 266, 713 

\bibitem[\protect\citeauthoryear{Pesce, Falomo, \& Treves}{1994}]{Pesce1994} Pesce J.~E., Falomo R., Treves A., 1994, AJ, 107, 494
\bibitem[\protect\citeauthoryear{Pesce, Falomo, \& Treves}{1995}]{Pesce1995} Pesce J.~E., Falomo R., Treves A., 1995, AJ, 110, 1554 

\bibitem[\protect\citeauthoryear{Prochaska et al.}{2011}]{Prochaska2011} Prochaska J.~X., Weiner B., Chen H.-W., Cooksey K.~L., Mulchaey J.~S., 2011, ApJS, 193, 28 
\bibitem[\protect\citeauthoryear{Richter et al.}{2011}]{Richter2011} Richter P., Krause F., Fechner C., Charlton J.~C., Murphy M.~T., 2011, A\&A, 528, A12
\bibitem[\protect\citeauthoryear{Rieger \& Volpe}{2010}]{Rieger2010} Rieger F.~M., Volpe F., 2010, A\&A, 520, A23

\bibitem[\protect\citeauthoryear{Sandrinelli, Covino, \& Treves}{2014a}]{Sandrinelli2014a} Sandrinelli A., Covino S., Treves A., 2014a, A\&A, 562, AA79 
\bibitem[\protect\citeauthoryear{Sandrinelli, Covino, \& Treves}{2014b}]{Sandrinelli2014b} Sandrinelli A., Covino S., Treves A., 2014b, ApJ, 793, LL1
\bibitem[\protect\citeauthoryear{Sbarufatti et al.}{2005}]{Sbarufatti2005a} Sbarufatti B., Treves A., Falomo R., Heidt J., Kotilainen J., Scarpa R., 2005, AJ, 129, 559 
\bibitem[\protect\citeauthoryear{Sbarufatti, Treves, \& Falomo}{2005}]{Sbarufatti2005b} Sbarufatti B., Treves A., Falomo R., 2005, ApJ, 635, 173
\bibitem[\protect\citeauthoryear{Sbarufatti et al.}{2006}]{Sbarufatti2006} Sbarufatti B., Falomo R., Treves A., Kotilainen J., 2006, A\&A, 457, 35 
\bibitem[\protect\citeauthoryear{Scarpa et al.}{1999}]{Scarpa1999} Scarpa R., Urry C.~M., Falomo R., Pesce J.~E., Webster R., O'Dowd M., Treves A., 1999, ApJ, 521, 134 
\bibitem[\protect\citeauthoryear{Scarpa et al.}{2000}]{Scarpa2000} Scarpa R., Urry C.~M., Falomo R., Pesce J.~E., Treves A., 2000, ApJ, 532, 740 
\bibitem[\protect\citeauthoryear{Schlafly et al.}{2014}]{Schlafly2014} Schlafly E.~F., et al., 2014, ApJ, 789, 15
\bibitem[\protect\citeauthoryear{Schwartz et al.}{1979}]{Schwartz1979} Schwartz D.~A., Griffiths R.~E., Schwarz J., Doxsey R.~E., Johnston M.~D., 1979, ApJ, 229, L53 
\bibitem[\protect\citeauthoryear{Sembay et al.}{1993}]{Sembay1993} Sembay S., Warwick R.~S., Urry C.~M., Sokoloski J., George I.~M., Makino F., Ohashi T., Tashiro M., 1993, ApJ, 404, 112 
\bibitem[\protect\citeauthoryear{Smith et al.}{1992}]{Smith1992} Smith P.~S., Hall P.~B., Allen R.~G., Sitko M.~L., 1992, ApJ, 400, 115
\bibitem[\protect\citeauthoryear{Smith, O'Dea, \& Baum}{1995}]{Smith1995} Smith E.~P., O'Dea C.~P., Baum S.~A., 1995, ApJ, 441, 113

\bibitem[\protect\citeauthoryear{Tody}{1986}]{Tody1986} Tody D., 1986, SPIE, 627, 733 
\bibitem[\protect\citeauthoryear{Tody}{1993}]{Tody1993} Tody D., 1993, ASPC, 52, 173 
\bibitem[\protect\citeauthoryear{Urry et al.}{1993}]{Urry1993} Urry C.~M., et al., 1993, ApJ, 411, 614 
\bibitem[\protect\citeauthoryear{Urry et al.}{2000}]{Urry2000} Urry C.~M., Scarpa R., O'Dowd M., Falomo R., Pesce J.~E., Treves A., 2000, ApJ, 532, 816 


\bibitem[\protect\citeauthoryear{van Dokkum}{2001}]{vanDokkum2001} van Dokkum P.~G., 2001, PASP, 113, 1420 
\bibitem[\protect\citeauthoryear{Weistrop, Smith, \& Reitsema}{1979}]{Weistrop1979} Weistrop D., Smith B.~A., Reitsema H.~J., 1979, ApJ, 233, 504
\bibitem[\protect\citeauthoryear{Worpel et al.}{2013}]{Worpel2013} Worpel H., Brown M.~J.~I., Jones D.~H., Floyd D.~J.~E., Beutler F., 2013, ApJ, 772, 64 
\bibitem[\protect\citeauthoryear{Wurtz et al.}{1993}]{Wurtz1993} Wurtz R., Ellingson E., Stocke J.~T., Yee H.~K.~C., 1993, AJ, 106, 869
\bibitem[\protect\citeauthoryear{Wurtz et al.}{1997}]{Wurtz1997} Wurtz R., Stocke J.~T., Ellingson E., Yee H.~K.~C., 1997, ApJ, 480, 547 
\bibitem[\protect\citeauthoryear{Yee \& L{\'o}pez-Cruz}{1999}]{Yee1999} Yee H.~K.~C., L{\'o}pez-Cruz O., 1999, AJ, 117, 1985 
\bibitem[\protect\citeauthoryear{Zacharias et al.}{2004}]{Zacharias2004} Zacharias N., Monet D.~G., Levine S.~E., Urban S.~E., Gaume R., Wycoff G.~L., 2004, AAS, 36, 1418 
\bibitem[\protect\citeauthoryear{Zacharias et al.}{2005}]{Zacharias2005} Zacharias N., Monet D.~G., Levine S.~E., Urban S.~E., Gaume R., Wycoff G.~L., 2005, yCat, 1297, 
\bibitem[\protect\citeauthoryear{Zhang et al.}{2014}]{Zhang2014} Zhang B.-K., Zhao X.-Y., Wang C.-X., Dai B.-Z., 2014, RAA, 14, 933 
\end{thebibliography}
\end{document}